\documentclass[format=acmsmall, review=false, screen=true, authorversion]{acmart}
\renewcommand\footnotetextcopyrightpermission[1]{}
\makeatletter
\renewcommand\@formatdoi[1]{\ignorespaces}
\makeatother
\usepackage{natbib}
\usepackage[utf8]{inputenc}
\usepackage{hyperref}
\usepackage[inline]{enumitem}
\usepackage[english]{babel}
\usepackage{colortbl}
\usepackage{csquotes}

\usepackage{footnote}
\makesavenoteenv{table}

\addto\extrasenglish{%

}

\usepackage[ruled]{algorithm2e} 

\SetAlFnt{\small}
\SetAlCapFnt{\small}
\SetAlCapNameFnt{\small}
\SetAlCapHSkip{0pt}
\IncMargin{-\parindent}

\presetkeys%
    {todonotes}%
    {inline,backgroundcolor=yellow}{}

\setlist[enumerate]{leftmargin=*}

\acmJournal{CSUR}
\acmVolume{54}
\acmNumber{4}
\acmArticle{105}
\acmYear{2021}
\acmMonth{5}
\copyrightyear{2021}

\setcopyright{acmlicensed}

\acmDOI{10.1145.3453154}

\definecolor{blue}{rgb}{0, 0, 0}

\begin{document}

\title{A Survey on Conversational Recommender Systems}

\author{Dietmar Jannach}
\affiliation{
\institution{University of Klagenfurt}
\city{Klagenfurt}
\country{Austria}
}
\author{Ahtsham Manzoor}
\affiliation{
\institution{University of Klagenfurt}
\city{Klagenfurt}
\country{Austria}
}
\author{Wanling Cai}
\affiliation{
\institution{Hong Kong Baptist University}
\city{Hong Kong}
\country{China}
}
\author{Li Chen}
\affiliation{
\institution{Hong Kong Baptist University}
\city{Hong Kong}
\country{China}
}

\begin{abstract}
Recommender systems are software applications that help users to find items of interest in situations of information overload.
Current research often assumes a one-shot interaction paradigm, where the users' preferences are estimated based on past observed behavior and where the presentation of a ranked list of suggestions is the main, one-directional form of user interaction. Conversational recommender systems (CRS) take a different approach and support a richer set of interactions. These interactions can, for example, help to improve the preference elicitation process or allow the user to ask questions about the recommendations and to give feedback. The interest in CRS has significantly increased in the past few years. This development is mainly due to the significant progress in the area of natural language processing, the emergence of new voice-controlled home assistants, and the increased use of chatbot technology. With this paper, we provide a detailed survey of existing approaches to conversational recommendation. We categorize these approaches in various dimensions, e.g., in terms of the supported user intents or the knowledge they use in the background. Moreover, we discuss technological approaches, review how CRS are evaluated, and finally identify a number of gaps that deserve more research in the future.

\end{abstract}



\thanks{Authors' addresses: Dietmar Jannach and Ahtsham Manzoor, University of Klagenfurt, Austria. Email: dietmar.jannach@aau.at, ahtsham.manzoor@aau.at. Wanling Cai and Li Chen, Hong Kong Baptist University, China. Email: cswlcai@comp.hkbu.edu.hk, lichen@comp.hkbu.edu.hk}

\maketitle
\renewcommand{\shortauthors}{Dietmar Jannach et al.}
\bibliographystyle{abbrvnat}

\section{Introduction}
\label{sec:introduction}
Recommender systems are among the most visible success stories of AI in practice. Typically, the main task of such systems is to point users to potential items of interest, e.g., in the context of an e-commerce site. Thereby, they not only help users in situations of information overload \cite{Ricci:2010:RSH:1941884}, but they also can significantly contribute to the business success of the service providers \cite{jannachjugovactmis2019}.

In many of these practical applications, recommending is a \emph{one-shot} interaction process. Typically, the underlying system monitors the behavior of its users over time and then presents a tailored set of recommendations in pre-defined navigational situations, e.g., when a user logs in to the service. 
Although such an approach is common and useful in various domains, it can have a number of potential limitations. There are, for example, a number of application scenarios, where the user preferences cannot be reliably estimated from their past interactions. This is often the case with high-involvement products (e.g., when recommending a smartphone), where we even might have no past observations at all. Furthermore, what to include in the set of recommendations can be highly context-dependent, and it might be difficult to automatically determine the user's current situation or needs. Finally, another assumption often is that users already know their preferences when they arrive at the site. This might, however, not necessarily be true. Users might also \emph{construct} their preferences only during the decision process \cite{ResearchNoteContingencyApproach2013}, when they become aware of the space of the options. 
In some cases, they might also learn about the domain and the available options only during the interaction with the recommender \cite{warnestaal2005user}.

The promise of Conversational Recommender Systems (CRS) is that they can help to address many of these challenges. The general idea of such systems, broadly speaking, is that they support a task-oriented, multi-turn dialogue with their users. During such a dialogue, the system can elicit the detailed and current preferences of the user, provide explanations for the item suggestions, or process feedback by users on the made suggestions. Given the significant potential of such systems, research on CRS already has some tradition. Already in the late 1970s, Rich \cite{RICH1979329} envisioned a computerized librarian that makes reading suggestions to users by interactively asking them questions, in \emph{natural language}, about their personality and preferences.
Besides interfaces based on natural language processing (NLP), a variety of \emph{form-based user interfaces}\footnote{With form-based UIs, we refer to systems approaches where users fill out forms and use check boxes or buttons.} were proposed over the years. One of the earlier interaction approaches in CRS based on such interfaces is called \emph{critiquing}, which was proposed as a means for query reformulation in the database field already in 1982 \cite{DBLP:conf/aaai/TouWFHM82}. In critiquing approaches, users are presented with a recommendation soon in the dialogue and can then apply pre-defined critiques on the recommendations, e.g., (``less \$\$'') \cite{hammond1994findme,DBLP:journals/expert/BurkeHY97}.

Form-based approaches can generally be attractive as the actions available to the users are pre-defined and non-ambiguous. However, such dialogues may also appear non-natural, and users might feel constrained in the ways they can express their preferences. NLP-based approaches, on the other hand, for a long time suffered from existing limitations, e.g., in the context of processing voice commands. In recent years, however, major advances were made in language technology. As a result, we are nowadays used to issuing voice commands to our smartphones and digital home assistants, and these devices have reached an impressive level of recognition accuracy.
In parallel to these developments in the area of voice assistants, we have observed a fast uptake of \emph{chatbot} technology in recent years. Chatbots, both rather simple and more sophisticated ones, are usually able to process natural language and are nowadays widely used in various application domains, e.g., to deal with customer service requests.

These technological advances led to an increased interest in CRS during the last years.
In contrast to many earlier approaches, we however observe that today's technical proposals are more often based on machine learning technology instead of following pre-defined dialogue paths. 
However, often there still remains a gap between the capabilities of today's voice assistants and chatbots compared to what is desirable to support truly conversational recommendation scenarios \cite{Rafailidis:2019:VAP:3326467.3326468}, in particular when the system is voice-controlled \cite{Yang:2018:UUI:3240323.3240389,wissbroecker2018early}.

In this paper, we review the literature on CRS   in terms of common building blocks of a typical conceptual architecture of CRS. Specifically,   after providing a definition and a conceptual architecture of a CRS in Section \ref{sec:definition}, we discuss
\begin{enumerate*}[label=\textit{(\roman*)}]
  \item interaction modalities of CRS (Section \ref{sec:interaction-modalities}),
  \item the knowledge and data they are based upon (Section \ref{sec:knowledge-behind}), and
  \item the computational tasks that have to be accomplished in a typical CRS (Section \ref{sec:computational-tasks}).
\end{enumerate*}
Afterwards, we discuss evaluation approaches for CRS (Section \ref{sec:evaluation}) and finally give an outlook on future directions.

\section{Definitions and Research Methodology}
\label{sec:definition}

In this section we discuss relevant preliminaries to our work. First, we provide a general characterization and conceptual model of CRS. Second, we discuss our research methodology.

\subsection{Characterization of Conversational Recommender Systems}
There is no widely-established definition in the literature of what represents a CRS. In this work, we use the following definition.

\begin{definition}[Conversational Recommender System--CRS] A CRS is a software system that supports its users in achieving recommendation-related goals through a multi-turn dialogue. 
\end{definition}

One fundamental characteristic of CRS is their task-orientation, i.e., they support recommendation specific tasks and goals. The main task of the system is to provide recommendations to the users, with the goal to support their users' decision-making process or to help them find relevant information. Additional tasks of CRS include the acquisition of user preferences or the provision of explanations. This specific task orientation distinguishes CRS from other dialogue-based systems, such as the early ELIZA system \cite{Weizenbaum:1966:ECP:365153.365168} or similar \emph{chat robot} systems \cite{WallaceALICE}.

The other main feature of a CRS according to our definition is that there is a \emph{multi-turn} conversational interaction. This stands in contrast to systems that merely support question answering (Q\&A tools). Providing one-shot Q\&A-style recommendations is a common feature of personal digital assistants like Apple's Siri and similar products. While these systems already today can reliably respond to recommendation requests, e.g., for a restaurant, they often face difficulties maintaining a multi-turn conversation. A CRS therefore explicitly or implicitly implements some form of \emph{dialogue state management} to keep track of the conversation history and the current state.

Note that our definition does not make any assumptions regarding the \emph{modality} of the inputs and the outputs. CRS can be voice controlled, accept typed text, or obtain their inputs via form fields, buttons, or even gestures. Likewise, the output is not constrained and can be voice, speech, text, or multimedia content. No assumptions are also made regarding who drives the dialogue. 

Generally, conversational recommendation shares a number of similarities with 
\emph{conversational search} \cite{Radlinski:2017:TFC:3020165.3020183}. 
In terms of the underlying tasks, search and recommendation have in common that one main task is to rank the objects according to their assumed relevance, either for a given query (search) or the preferences of the user (recommendation). Furthermore, in terms of the conversational part, both types of systems have to interpret user utterances and disambiguate user intents in case natural language interactions are supported. In conversational search systems, however, the assumption often is that the interaction is based on ``written or spoken form'' \cite{Radlinski:2017:TFC:3020165.3020183}, whereas in our definition of CRS various types of input modalities are possible.
Overall, the boundary between (personalized) conversational search and recommendation systems often seems blurry, see \cite{Zhang:2018:TCS:3269206.3271776,Sun:2018:CRS:3209978.3210002,Mahmood:2009:IRS:1557914.1557930}, in particular as often similar technological approaches are applied. In this survey, we limit ourselves to works that explicitly mention recommendation as one of their target problems.

\subsection{Conceptual Architecture of a CRS}
\begin{figure}[h!t]
  \centering
  \includegraphics[width=0.85\linewidth]{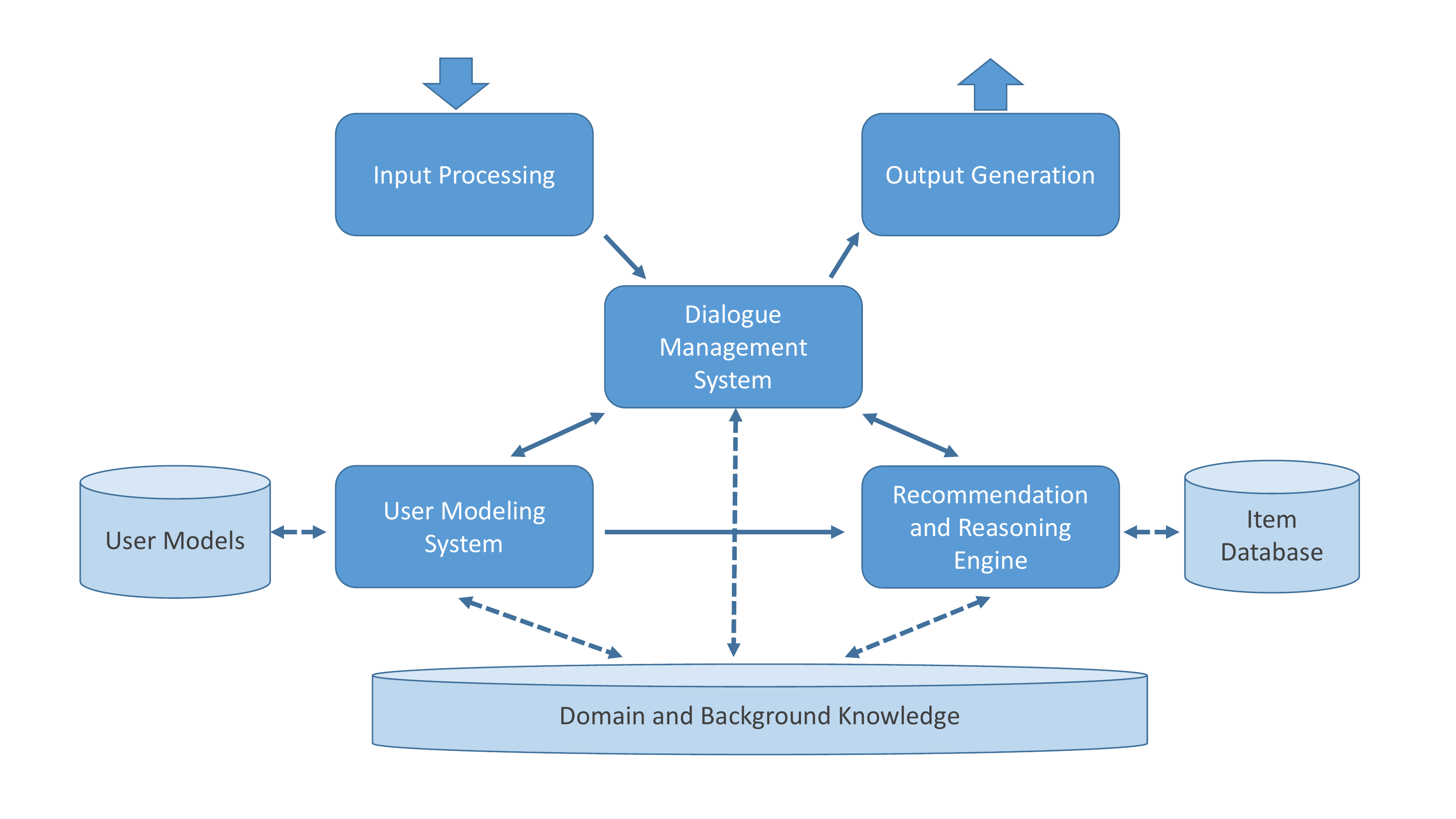}
  \caption{Typical architecture of a conversational recommender system (see also \cite{Thompson:2004:PSC:1622467.1622479}).}\label{fig:architecture}
\end{figure}

A variety of technical approaches for building CRS were proposed in the last two decades. The specifics of the technical architecture of such solutions depend on the system's functionality, i.e., whether or not voice input is supported. Still, a number of typical interoperating conceptual components of such architectures can be identified, as shown in Figure \ref{fig:architecture}.

\paragraph{Computational Elements}
One central part of such an architecture usually is a \emph{Dialogue Management System} (also called ``state tracker'' or similarly in some systems).   This component drives the process flow. It receives the processed inputs, e.g., the recognized intents, entities and preferences, and correspondingly updates the dialogue state and user model. After that, using a recommendation and reasoning engine and background knowledge, it determines the next action and returns appropriate content like a recommendation list, an explanation, or a question to the output generation component.

The \emph{User Modeling System} can be a component of its own, in particular when there are long-term user preferences to be considered, or not. In some cases, the current preference profile is implicitly part of the dialogue system. The \emph{Recommendation and Reasoning Engine} is responsible for retrieving a set of recommendations, given the current dialogue state and preference model. This component might also implement other complex reasoning functionality, e.g., to generate explanations or to compute a query relaxation (see later). Besides these central components, typical CRS architectures comprise modules for input and output processing. These can, for example, include speech-to-text conversion and speech generation. On the input side---in particular in the case of natural language input---additional tasks are usually supported, including \emph{intent detection} and \emph{named entity recognition} { \cite{Kapetanios:2013NLP,nadeau2007survey}, for identifying the users' intentions and entities (e.g., attributes of item) in their utterances.}

\paragraph{Knowledge Elements}
Various types of knowledge are used in CRS. The \emph{Item Database} is something that is present in almost all solutions, representing the set of recommendable items, sometimes including details about their attributes. In addition to that, different types of \emph{Domain and Background Knowledge} are often leveraged by CRS. Many approaches explicitly encode \emph{dialogue knowledge} in different ways, e.g., in the form of pre-defined dialogue states, supported user intents, and the possible transitions between the states. This knowledge can be general or specific to a particular domain. The knowledge can furthermore either be encoded by the system designers or automatically learned from other sources or previous interactions. A typical example for learning approaches are those that use machine learning to build statistical models from corpora of recorded dialogues.
Generally, domain and background knowledge can be used by all computational elements. Input processing may need information about entities to be recognized or knowledge about the pre-defined intents. The user modeling component may be built on estimated interest weights regarding certain item features, and the reasoning engine may use explicit inference knowledge to derive the set of suitable recommendations.

\subsection{Research Method: Identifying Relevant Works}
We followed a semi-systematic approach to identify relevant papers. We first queried several digital libraries\footnote{We looked at Springer Link, the ACM Digital Library, IEEE Xplore, ScienceDirect, arXiv.org, and ResearchGate} using pre-defined search strings such as
``conversational recommender system'', ``interactive recommendation'', ``advisory system'', or ``chatbot recommender''. The returned papers were then manually checked for their relevance based on titles and abstracts. Papers considered relevant were read in detail and, if considered to be in the scope of the paper, used as a starting point for a snowballing procedure.
 Overall, the paper selection process surfaced 121 papers on CRS that we considered in
 this work.\footnote{A few additional papers were considered later on based on reviewer suggestions. Overall, while this research is not the result of a \emph{systematic} literature review, we are confident that the selection of considered papers is not biased, given our structured and documented search process.\color{black}} Looking at the type of these papers, the majority of the works described technical proposals for one of the computational components of a CRS architecture. A smaller set of papers described demo systems. Another smaller set were analytical ones which, for example, reviewed certain general characteristics of CRS.

Generally, we only included papers that are compliant with our definition of a CRS given above.
We therefore did not include papers that discussed one-shot or multi-step question-answering systems \cite{Yin:2017:DID:3097983.3098148,SinghcinICD102019}, even when the question or task was about a recommendation. We also did not consider general dialogue systems like chatbot systems, which are not task-oriented, or systems that only support a query-response interaction process  like a search engine without further dialogue steps, e.g., \cite{Clarizia2018}. Furthermore, we did not include dialogue-based systems, which were task-oriented, but not on a recommendation task, e.g., the end-to-end learning approaches presented in \cite{wen-etal-2017-network} and \cite{li2017end-to-end}, which focus on restaurant search and movie-ticket booking. 
Furthermore, we excluded a few works like \cite{Hariri:2014:CAI:2645710.2645753} or \cite{Zhao:2013:ICF:2541154.2505690}, which use the term ``interactive recommendation'', which however rather refers to a system that addresses observed user interest changes over time, but is not designed to support a dialogue with the user. 
Other works like \cite{Sun:2013:LMD:2433396.2433451} or \cite{Zhao:2013:ICF:2541154.2505690} mainly focus on finding good strategies for acquiring an initial set of ratings for cold-start users. While these works can be seen as supporting an interactive process, there is only one type of interaction, which is furthermore mostly limited to a profile-building phase. Finally, there are a number of works where users of a recommendation systems are provided with mechanisms to fine-tune their recommendations, which is sometimes referred to as ``user control'' \cite{JannachNaveedEtAl2016}.
Such works, e.g., \cite{DBLP:journals/jmis/XuBC17}, in principle support user actions that can be found in some CRS, for example to give feedback on a recommendation. The interaction style of such approaches is however not a dialogue with the system.

\section{Interaction Modalities of CRS}
\label{sec:interaction-modalities}

The recent interest in CRS is spurred both by developments in NLP and technological advances such as broadband mobile internet access and new devices like smartphones and home assistants. Our review of the literature however shows that the interaction between users and a CRS is neither limited to natural language input and output nor to specific devices.

\subsection{Input and Output Modalities}
The majority of the surveyed papers explicitly or implicitly support two main forms of inputs and outputs, either as the only modality or combined in a hybrid approach:
\begin{itemize}
\item Based on forms and structured layouts, as in a traditional web-based (desktop) application.
\item Based on natural language, either in written or spoken form.
\end{itemize}

Approaches that are \emph{exclusively} based on forms (including buttons, radio-buttons etc.) and structured text for the output are common for almost all (except \cite{Grasch:2013:RTC:2507157.2507161})
critiquing-based approaches, e.g.,
\cite{AverjanovaMobyRec2008,DBLP:journals/expert/RicciN07,Hong:2010:IAS:2108616.2108681,Dietz2019,McCarthy2004Onthe}, as well as for web-based interactive advisory solutions as presented, e.g., in \cite{Felfernig:2006:IED:1278152.1278154,Jannach:2004:ASK:3000001.3000153,JannachENTER2007}. In such applications, users typically follow pre-defined dialogue paths and interact with the application by filling forms or choosing from pre-defined options. The final output typically is a structured visual representation in the form of a list of options.

On the other hand, approaches that are \emph{entirely} based on natural language interactions include task-oriented dialogue systems like the early proposal from \cite{RICH1979329}, the explanation-aware conversational system proposed in \cite{Pecune:HAI2019}, as well as more recent (deep) learning-based approaches, e.g., \cite{Greco2017,liao2019deep,DBLP:conf/aaai/GhazvininejadBC18}. \emph{Spoken-text-only} approaches are often implemented on smart speakers like Amazon Alexa or Google Home, e.g., \cite{Dalton:2018:VGC:3209978.3210168,Argal2018}. Compared to form-based approaches, these solutions usually offer more flexibility in the dialogue and sometimes support chit-chat and mixed-initiative dialogues. Major challenges can, however, lie in the understanding of the users' utterances and the identification of their intents. But also the presentation of the recommendations can be difficult, in particular when more than one option should be provided at once.

\emph{Hybrid} approaches that combine natural language with other modalities are, therefore, not uncommon. For example, systems that support written natural language dialogues often rely on list-based or other visual approaches to present their results \cite{Zhang:2018:TCS:3269206.3271776,DBLP:conf/aaai/LeeMRAJ18}. The work presented in \cite{Yu:2019:VDA:3292500.3330991}, on the other hand, supports a hybrid visual/natural language interaction mechanism, where recommendations are displayed visually, and users can provide feedback  to certain features in a critiquing-like form in natural language. Yet other systems support voice input, but present the recommendations in textual form \cite{Grasch:2013:RTC:2507157.2507161,Thompson:2004:PSC:1622467.1622479}, because it can be difficult to present more than one recommendation at a time through spoken language without overwhelming the users. \emph{Chatbot} applications, finally, often combine natural language input and output with structured form elements (e.g., buttons) and a visually-structured representation of the recommendations \cite{qiu2017alime,Jin:2019:MEC:3357384.3357923,DBLP:conf/recsys/NarducciBIGLS18,Ikemoto2019}.

Besides written or spoken language and fill-out forms, a few other alternative and \emph{application-specific modalities} for inputs and outputs can be found. The dialogue system presented in \cite{WALKER2004811}, for example, supports multiple types of inputs, including \emph{visual inputs} on a geographic map, \emph{pen gestures} like zooming, or \emph{handwritten input}. The work proposed in \cite{DECAROLIS201787} furthermore tries to process \emph{non-verbal} input, like \emph{body postures}, \emph{gestures}, \emph{facial expressions}, as well as \emph{speech prosody} to estimate the user's emotions and attitudes in order to acquire implicit feedback and preferences.

In terms of the \emph{outputs}, several approaches use interactive geographic \emph{maps}, often as part of a multi-modal output strategy \cite{WALKER2004811,DBLP:conf/aaai/LeeMRAJ18,Dietz2019,Arteag2019}. The applicability of map-based approaches is  limited to certain application domains, e.g., travel and tourism, but can help to overcome various challenges regarding the user experience with conversational systems \cite{MapBasedRicci2010}.  The use of \emph{embodied conversational agents} (ECAs) \cite{Cassell:2001:ECA:567363.567368} as an additional output mechanism is also not uncommon in the literature \cite{Felfernig:2006:IED:1278152.1278154,Hong:2010:IAS:2108616.2108681} because of the assumed general persuasive potential of  human-like avatars \cite{DEHN20001,Andre2010}. Various factors can impact the effectiveness of such ECAs. In \cite{Foster2010umuai}, for example, the authors analyze the effects of \emph{non-verbal} behavior (e.g., the facial expressions) on the effectiveness of an ECA in the context of a dialogue-based recommender system. Research on the specific effects of using different variants of an ECA in the context of recommender systems is, however, generally rare.

\begin{figure}[h!t]
  \centering
  \includegraphics[width=0.95\linewidth]{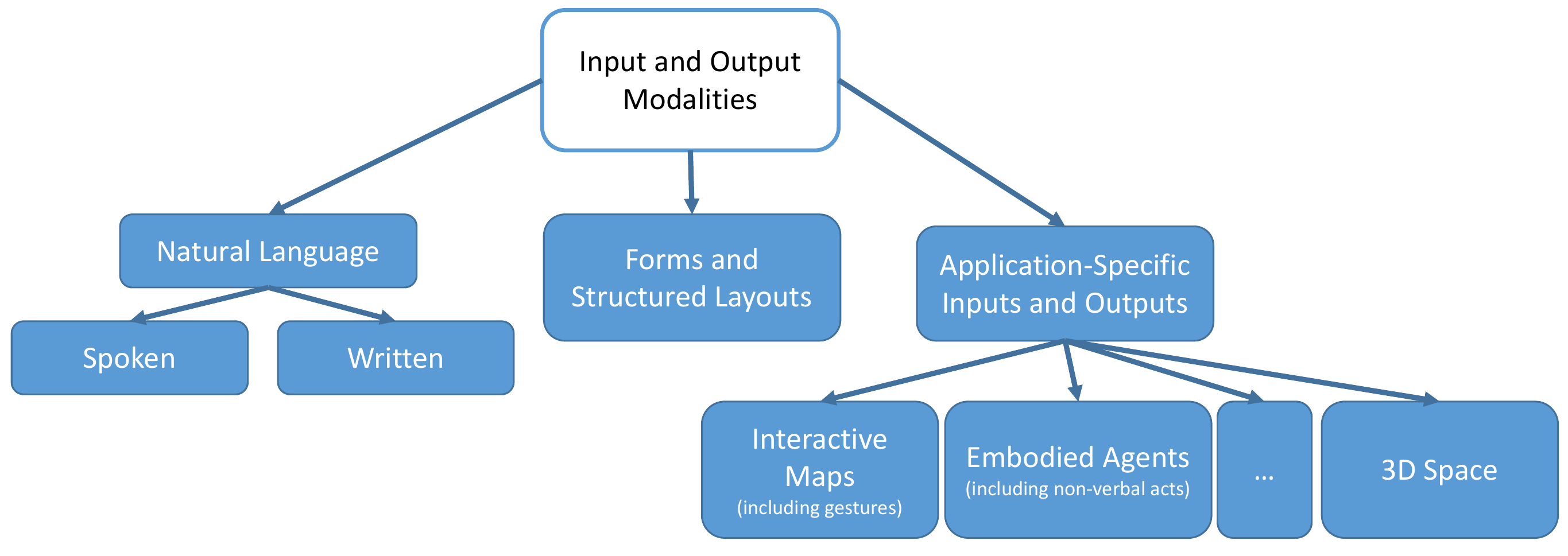}
  \caption{Categorization of input and output modalities.}\label{fig:modalities}
\end{figure}

Finally, a few works exist where users interact with a recommendation system within a \emph{virtual, three-dimensional space}. In \cite{Contreras2018,Contreras2015A3V}, the authors describe a virtual shopping environment where users interact with a critiquing-based recommender and can, in addition, collaborate with other users. Supporting group decisions is also the goal of the work presented in \cite{Alvarez2016}. In this work, however no 3D visualization is supported, and the focus of the work is mostly to enable the conversation between a group of users supported by a recommender system. Figure \ref{fig:modalities} provides an overview of common input and output modalities found in the literature.

\subsection{Application Environment}
\paragraph{Stand-alone and Embedded Applications}
CRS can both be stand-alone applications or part of a larger software solution. In the first case, recommendation is the central functionality of the system. Examples for such applications include the mobile tourist guides proposed in \cite{Mahmood:2009:IRS:1557914.1557930,AverjanovaMobyRec2008,JannachENTER2007}, the interactive e-commerce advisory systems discussed in \cite{Felfernig:2006:IED:1278152.1278154,Jannach:2007:RDK:2011243.2011249}, or the early \emph{FindMe} browsing and shopping systems \cite{Burke:1999:WPS:315149.315486,DBLP:journals/expert/BurkeHY97}.
In the second case, that of an embedded application, the CRS does not (entirely) stand on its own.
Often, the CRS is implemented in the form of chatbot that is embedded within e-commerce solutions \cite{BotWheels2017,DBLP:conf/aaai/YanDCZZL17}
or other types of web-portals \cite{FuzzyBased2018}. In some cases, the CRS is also part of a multi-modal 2D or 3D user experience, like in \cite{Contreras2015A3V} and \cite{Foster2010umuai}. A special case in this context is the use of a CRS on voice-based home assistants (smart speakers) \cite{Argal2018,Dalton:2018:VGC:3209978.3210168}. In such settings, providing recommendations is only one of many functionalities the device is capable of. Users might therefore not actually perceive the system as primarily being a recommender.

\paragraph{Supported Devices} An orthogonal aspect regarding the application environment of a CRS is that of the supported devices. This is particularly important, because the specific capabilities and features of the target device can have a significant impact on the design choices when building a CRS. The mentioned smart speaker applications, for example, are specifically designed for hardware devices that often only support voice-based interactions. This can lead to specific challenges, e.g., when it comes to determining the user's intent or when a larger set of alternatives should be presented to the users. The interaction with chatbot applications, on the other hand, is typically not tied to specific hardware devices. Commonly, they are either designed as web applications or as smartphone and tablet applications. However, the choice of the used communication modality can still depend on the device characteristics. Typing on small smartphone screens may be tedious and the limited screen space in general requires the development of tailored user interfaces.

The applicability of CRS is not limited to the mentioned devices. Alternative approaches were, for example, investigated in \cite{DECAROLIS201787,decarolis2015empire}. Here, the idea is that the CRS is implemented as an application on an interactive wall that could be installed in a real store. A camera is furthermore used to monitor and interpret the user's non-verbal communication actions, in particular facial expressions and gestures. An alternative on-site environment was envisioned in \cite{zeng2018eliciting}. Here, the ultimate goal is to build a CRS running on a service robot, in this case one that is able to elicit a customer's food preferences in a restaurant. Yet another application scenario, that of future in-car recommender systems, is sketched in \citep{Luettin2019}. Given the specific situation in a driving scenario, the use of speech technology often is  advisable \citep{Chen2010}, which almost naturally leads to conversational recommendation approaches, e.g., for driving-related aspects like navigation or entertainment \citep{Bader2011,Becker2006PAIS}.

\subsection{Interaction Initiative}
A central design question for most conversational systems is who takes the initiative in the dialogue. Traditionally, we can differentiate between
\begin{enumerate*}[label=\textit{(\roman*)}]
  \item system-driven,
  \item user-driven, and
  \item mixed-initiative systems.
\end{enumerate*}
When considering CRS primarily as dialogue systems, such a classification can in principle be applied as well, but the categorization is not always entirely clear.

Critiquing-based systems are often considered to be mainly \emph{system-driven}, and sometimes mixed-initiative, e.g., in \cite{DBLP:conf/recsys/ViappianiPF07}. In such applications, the users are typically first asked about their preferences, e.g., using a form, and then an initial recommendation is presented. Users can then use a set of pre-defined or dynamically determined critiques to further refine their preferences. While the users in such applications have some choices regarding the dialogue flow, e.g., they can decide to accept a recommendation or further apply critiques, these choices are typically very limited and the available critiques are determined by the system.
Another class of mostly system-driven applications are the form-based interactive advisory systems discussed in \cite{Felfernig:2006:IED:1278152.1278154}. Here, the system guides the user through a personalized preference elicitation dialogue until enough is known about the user. Only after the initial recommendations are displayed, the user can influence the dialogue by selecting from pre-defined options like asking for an explanation or by relaxing some constraints.

The other extreme would be a \emph{user-driven} system, where the system takes no proactive role. The resulting dialogue therefore consists of ``user-asks, system-responds'' pairs, and it stands to question if we would call such an exchange a conversational recommendation. Such conversation patterns are rather typical for one-shot query-answering, search and recommendation systems that are not in the scope of our survey. As a result, in the papers considered relevant for this study, we did not find any paper that aimed at building an \emph{entirely} user-driven system in which the system never actively engages in a dialogue, e.g., when it does not ask any questions ever. A special case in that context is the recommender system proposed in \cite{Stanley2010}, which monitors an ongoing group chat and occasionally makes recommendations to the group based on the observed communication.

This observation is not surprising because every CRS is a task-oriented system aiming to achieve goals like obtaining enough reliable information about the user's preferences.  As a result, almost all approaches in the literature are \emph{mixed-initiative} systems, although with different degrees of system guidance. Typical chatbot applications, for example, often guide users through a series of questions with pre-defined answer options (using forms and buttons), and at the same time allow them to type in statements in natural language. In fully NLP-based interfaces, users typically have even more freedom to influence how the dialogue continues. Still, also in these cases, the system typically has some agenda to move the conversation forward.

Technically, even a fully NLP-based dialogue can almost entirely be system-driven and mostly rely on a ``system asks, user responds'' \citep{Zhang:2018:TCS:3269206.3271776} conversation pattern. Nonetheless, the provision of a natural language user interface might leave the users disappointed when they find out that they can never actively engage in the conversation, e.g., by asking a clarification question or explanation regarding the system's question.

\subsection{Discussion}
A variety of ways exist in which the user's interaction with a CRS can be designed, e.g., in terms of the input and output modalities, the supported devices, or the level of user control. In most surveyed papers, these design choices are, however, rarely discussed. One reason is that in many cases the proposed technical approach is mostly independent of the interaction modality, e.g., when the work is on a new strategy to determine the next question to ask to the user. In other cases, the modalities are pre-determined by the given research question, e.g., how to build a CRS on a mobile.

More research therefore seems required to understand how to make good design choices in these respects and what the implications and limitations of each design choice are. Regarding the chosen form of inputs and outputs, it is, for example, not always entirely clear if natural language interaction makes the recommendation more efficient or effective compared to form-based inputs. Pure natural language interfaces in principle provide the opportunity to elicit preferences in a more natural way. However, these interfaces have their limitations as well. The accuracy of the speech recognizer, for example, can have a major impact on the system's usability. In addition, some users might also be better acquainted and feel more comfortable with more traditional interaction mechanisms (forms and buttons). According to the study in \cite{iovine2020conversational}, a mix of a natural language interface and buttons led to the best user experience.  Moreover, in \cite{narducci2019investigation}, it turned out that in situations of disambiguation, i.e., when a user has to choose among a set of multiple alternatives, mixed-interaction mode (NLP interface with buttons) can make the task easier for users.
Overall, while in some cases the choice of the modalities is predetermined through the device, finding an optimal combination of interaction modalities remains challenging, in particular as individual user preferences might play a role here.

More studies are also needed to understand how much flexibility in the dialogue is required by users or how much active guidance by the system is appreciated in a certain application. Furthermore, even though language-based and in particular voice-based conversations have become more popular in recent years, certain limitations remain.
It is, for example, not always clear how one would describe a set of recommendations when using voice output. Reading out more than one recommendation seems impractical in most cases and something that we could call ``recommendation summarization'' might be needed.

Despite these potential current limitations, we expect a number of new opportunities where CRS can be applied in the future. With the ongoing technological developments, more and more devices and machines are equipped with CPUs and are connected to the internet. In-store interactive walls, service robots and in-car recommenders, as discussed above, are examples of visions that are already pursued today. These new applications will, however, also come with their own general challenges (e.g., privacy considerations, aspects of technology acceptance) and application-specific ones (e.g., safety considerations in an in-car setting).

\section{Underlying Knowledge and Data}
\label{sec:knowledge-behind}
Depending on the chosen technical approach, CRS have to incorporate various types of knowledge and background data to function. Clearly, like any recommender, there has to be information about the recommendable items. Likewise, the generation of the recommendations is either based on explicit knowledge, for example recommendation rules or constraints, or on machine learning models that are trained on some background data. However, conversational systems usually rely on additional types of knowledge about  the user intents that the CRS supports, the possible states in the dialogue, or data such as recorded and transcribed natural language recommendation dialogues that are used to train a machine learning model. In the following sections, we provide an overview on the different types of knowledge and data that were used in the literature to build a CRS.

\subsection{User Intents}
\label{subsec:supported-user-intents}
CRS are dialogue systems designed to serve very specific
purposes in the context of information filtering and decision making. Therefore, they have to support their users' particular information needs and intents that can occur in such conversations. In many CRS, the set of user intents that the system supports is pre-defined and represents a major part of the manually engineered background knowledge on which the system is built. In particular in NLP-based approaches, detecting the current user's intent and selecting the system's response is one of the main computational tasks of the system, see also Section \ref{sec:computational-tasks}.
In this section, we will therefore mainly focus on NLP-based systems.

The set of user intents that the system supports varies across the different CRS that are found in the literature, and the choice of which intents to support ultimately depends on the requirements of the application domain.
However, while a subset of the intents that the system supports is sometimes specific to the application as well, there are a number of intents that are common in many CRS. In Table \ref{tab:user-intents}, we provide a high-level overview of \emph{domain-independent} user intents that we have found in our literature review. The order of the intents in Table \ref{tab:user-intents} roughly follows the flow of a typical recommendation dialogue.  This overview is also intended to serve as a tool for the designers of CRS to check if there are any gaps in their current system with respect to potential user needs that are not well supported.

Research on what are relevant user intents is generally scarce, and we only found 11 papers that explicitly discussed user intents. Among these 11, only few of them, e.g., see \cite{cai2019department,DBLP:conf/recsys/NarducciBIGLS18,DBLP:conf/aaai/YanDCZZL17}, considered the majority of the domain-independent intents shown in Table \ref{tab:user-intents}. Others like  \cite{Thompson:2004:PSC:1622467.1622479,nie2019multimodal,kang2017understanding} only discuss certain subsets of them. Yet another set of papers focused on very application-specific intents in the context of group recommendation \cite{nguyen2017chat,Alvarez2016}.

\begin{table}[h!t]
\caption{\label{tab:user-intents} High-level overview of selected domain-independent user intents found in the literature.}
\small
\begin{tabular}{lp{9cm}}
\hline
\textbf{Intent Name}                                  &   \textbf{Intent Description} \\ \hline
\emph{Initiate Conversation}                          &  Start a dialogue with the system. \\
\emph{Chit-chat}                                      &  Utterances unrelated to the recommendation goal.  \\
\emph{Provide Preferences}                            &  Share preferences with the system.  \\
\emph{Revise Preferences}                             &  Revise previously stated preferences.  \\
\emph{Ask for Recommendation}                         &  Obtain system suggestions.  \\
\emph{Obtain Explanation}                             &  Learn more about why something was recommended.  \\
\emph{Obtain Details}                                 &  Ask about more details of a recommended object.  \\
\emph{Feedback on Recommendation}                     &  Give feedback on the provided recommendation(s).  \\
\emph{Restart}                                        &  Restart the dialogue.\\
\emph{Accept Recommendation}                          &  Accept one of the recommendations.\\
\emph{Quit}                                           &  Terminate the conversation.
                                               \\ \hline
\end{tabular}
\end{table}

\paragraph{Starting, re-starting, and ending the dialogue.} In NLP-based CRS, either the system or the user can initiate the dialogue. In a user-driven conversation, the recommendation seeker might, for example, explicitly ask for help \citep{DBLP:conf/recsys/NarducciBIGLS18} or make a recommendation request \citep{wu2019transferable} to start the interaction. One typical difficulty in this context is to recognize such requests when the dialogue starts with chit-chat. Once the recommendation dialogue is moving on, it is not uncommon that users want to start over, i.e., begin the session from scratch and ``reset their profile'' \citep{DBLP:conf/recsys/NarducciBIGLS18}. Previous studies found that such an intent was found in 5.2\% of the dialogues \citep{cai2019department} or that 36.4\% of the users had this intent in a conversation \citep{kang2017understanding}. Finally, at the end of the conversation, the user has either found a recommendation useful and accepts it in some form (e.g., by purchasing or consuming an item) or not. In either case, the CRS has to react to the intent in some form by redirecting the user accordingly, e.g., to the shopping basket, or by saying goodbye.

\paragraph{Chit-chat.} Many NLP-based systems support chit-chat in the conversation. In the study in \cite{DBLP:conf/aaai/YanDCZZL17}, nearly 80\% of the recorded user utterances were considered chit-chat. This number indicates that supporting chit-chat conversations can be a valuable means to create an engaging user experience. Furthermore, the study in \cite{DBLP:conf/aaai/YanDCZZL17} showed that chit-chat can also help to reduce user dissatisfaction, even though this part of the conversation is irrelevant to achieving the interaction goal.

\paragraph{Preference Elicitation.} Understanding the user's preferences is a key task for any CRS. Preference information can be provided by the user in different ways. In an initial phase of the dialogue, the user might specify some of the desired characteristics of the item that she or he is interested in or even provide strict filtering constraints.
In \citep{nie2019multimodal}, this process is termed as ``give criteria''. In later phases, the user might however also want to revise the previously stated preferences. 
Note that some authors also consider \emph{answering}---to a system-provided question or proposal for a constraint \citep{DBLP:conf/ictai/TrabelsiWBR10,chen2007preference}---as a dialogue intent during preference elicitation \citep{warnestaal2007pcql}. Since in NLP-based systems a user may respond in an arbitrary way,
it is clearly important for the system to disambiguate an answer by the user from other utterances. Such an ``Answer'' intent nonetheless is different from the other intents discussed here, as the intent is a response to the system's initiative of asking.

Also later in the process, preferences can be stated by the user in different ways after an initial recommendation is made by the system. In critiquing-based approaches, the users can, for example, add additional constraints in case the choice set is too large, relax some of the previously stated ones, or state that they already know the item \citep{Thompson:2004:PSC:1622467.1622479,Jannach2006,DBLP:journals/expert/RicciN07}. Generally, a system might also allow the user to inspect, modify, and delete the current profile (supporting a ``show profile'' intent) \citep{DBLP:conf/recsys/NarducciBIGLS18}. By analyzing the interaction logs of a prototypical voice-controlled movie recommender, e.g., in \cite{kang2017understanding}, the authors found that many users (41.1\%) at some stage try to refine their initially stated preferences. In particular in case of unsatisfactory system responses, some users might furthermore also have the intent to ``reject'' \citep{warnestaal2007pcql} a recommendation or ``restate'' their preferences. In the study presented in \cite{cai2019department}, this however happened only in 1.5\% of the interactions.
\paragraph{Obtaining Recommendations and Explanations.}
There are various ways in which users might ask for recommendations and additional information about the items. Asking for recommendations often happens at the very beginning of a dialogue, but this event can also occur after the user has revised the preferences. In case a currently displayed list of options is not satisfactory, users also might ask the system to ``show more'' options \citep{cai2019department} or ask for a similar item for comparison. For each of the items, the user might want to learn more about its details or ask for an explanation, e.g., why it was recommended \citep{DBLP:conf/recsys/NarducciBIGLS18}. Finally, an alternative form of requesting a recommendation is to ask the system about its opinion (``how about'') regarding a certain item, see e.g., \citep{DBLP:conf/aaai/YanDCZZL17}.

\subsection{User Modeling}
The interactive elicitation of the user's \emph{current} preferences or needs and constraints is another central task of CRS. As discussed above, this can be done through different modalities and by supporting various ways for users to express what they want. The acquired preferences are typically recorded in explicit form within a \emph{user profile}, based on which further inferences regarding the relevance of individual items can be made.
 There are two main ways of representing the preference information in such explicit models:
\begin{itemize}
  \item Preference expressions or estimates regarding \emph{individual items}, e.g., ratings, like and dislike statements, or implicit feedback signals. In \cite{Dietz2019}, for example, users are initially presented with a number of tourism destinations and asked which of them match their preferences.
  \item Preferences regarding \emph{individual item facets}. These facets can either relate to item attributes (e.g., the genre of a movie) or the desired functionalities.  \end{itemize}

For the latter class of approaches, the goal of the CRS is sometimes referred to as ``slot filling'', i.e., the recommender seeks to obtain values for a set of pre-defined facets. Sometimes, also the preference \emph{strength}, e.g., must or wish, can be relevant \cite{DBLP:journals/expert/RicciN07}. While different approaches for mining possible facets from structured and unstructured text documents were proposed in the literature \citep{WEI201580}, the set of facets is often manually engineered based on domain expertise. Furthermore, in case the facets refer to functional requirements as in \cite{widyantoro2014framework,Jannach:2004:ASK:3000001.3000153}, additional knowledge has to be encoded in the system to match these requirements with the recommendable items. In \citep{Jannach:2004:ASK:3000001.3000153}, for example, the user of an interactive camera recommender is asked about the type of photos she or he wants to take (e.g., sports photography), and a constraint-based system is then used to determine cameras that are suited for this purpose.

Finally, a few works exist that do \emph{not} assume the existence of a set of engineered item features. In \cite{ArbpitNavigationByPreference2020}, for example, an approach is proposed for preference elicitation, where user repeatedly specify preferences on items and the system then finds items that are similar in terms of unstructured features like keywords or tags. Similarly, other types of non-engineered features (tags, key phrases, or latent representations) were used in the preference elicitation approaches proposed in  \cite{LuoLatentLinearCritquing20,DBLP:conf/chi/LoeppHZ14} and \cite{vigTagGenome2011}.

Besides such ephemeral user models that are constructed during the ongoing session, some approaches in the literature also maintain long-term preference profiles \citep{DBLP:journals/expert/RicciN07,Thompson:2004:PSC:1622467.1622479,Argal2018,warnestaal2005user}. In the critiquing approach in \cite{DBLP:journals/expert/RicciN07}, for example, the system tries to derive long-term and supposedly more stable preferences (e.g., for  non-smoking rooms in restaurants) from multiple sessions. In the content-based recommendation approach adopted in \cite{Thompson:2004:PSC:1622467.1622479}, a probabilistic model is maintained based on past user preferences for items. In general, a key problem when recommending based on two types of models (long-term and short-term) is to determine the relative importance of the individual models.   One so far unexplored option could lie in the consideration of contextual factors such as seasonal aspects, the user's location, or the time of the day.

Finally, there are also approaches that try to leverage information about the collective preferences of a user community, in particular for cold-start situations \cite{Narducci2018aiai}. If nothing or little is known yet about the user's preferences, a common strategy is to recommend popular items, where item popularity can be determined based on user ratings, reviews, or past sales numbers as in \cite{Grasch:2013:RTC:2507157.2507161}. The feedback obtained for these popular items can then be used to further refine the user model.

\subsection{Dialogue States}
To be able to support a multi-turn, goal-oriented dialogue with their users, CRS have to implement appropriate means to keep track of the state of the dialogue to decide on the next \emph{conversational move}, i.e., the next action.
In many CRS implementations and in particular in knowledge-based approaches, dialogue management is based on a finite state machine, which not only defines the possible states but also the allowed transitions between the states \cite{Mahmood:2009:IRS:1557914.1557930, warnestaal2007interview, warnestaal2005modeling, mahmood2007learning}. In the Advisor Suite framework \citep{Jannach:2007:RDK:2011243.2011249,Jannach:2004:ASK:3000001.3000153}, for example, the entire recommendation logic including the dialogue flow was modelled with the help of graphical editors.

\begin{figure}[h!t]
  \centering
  \includegraphics[width=0.65\linewidth]{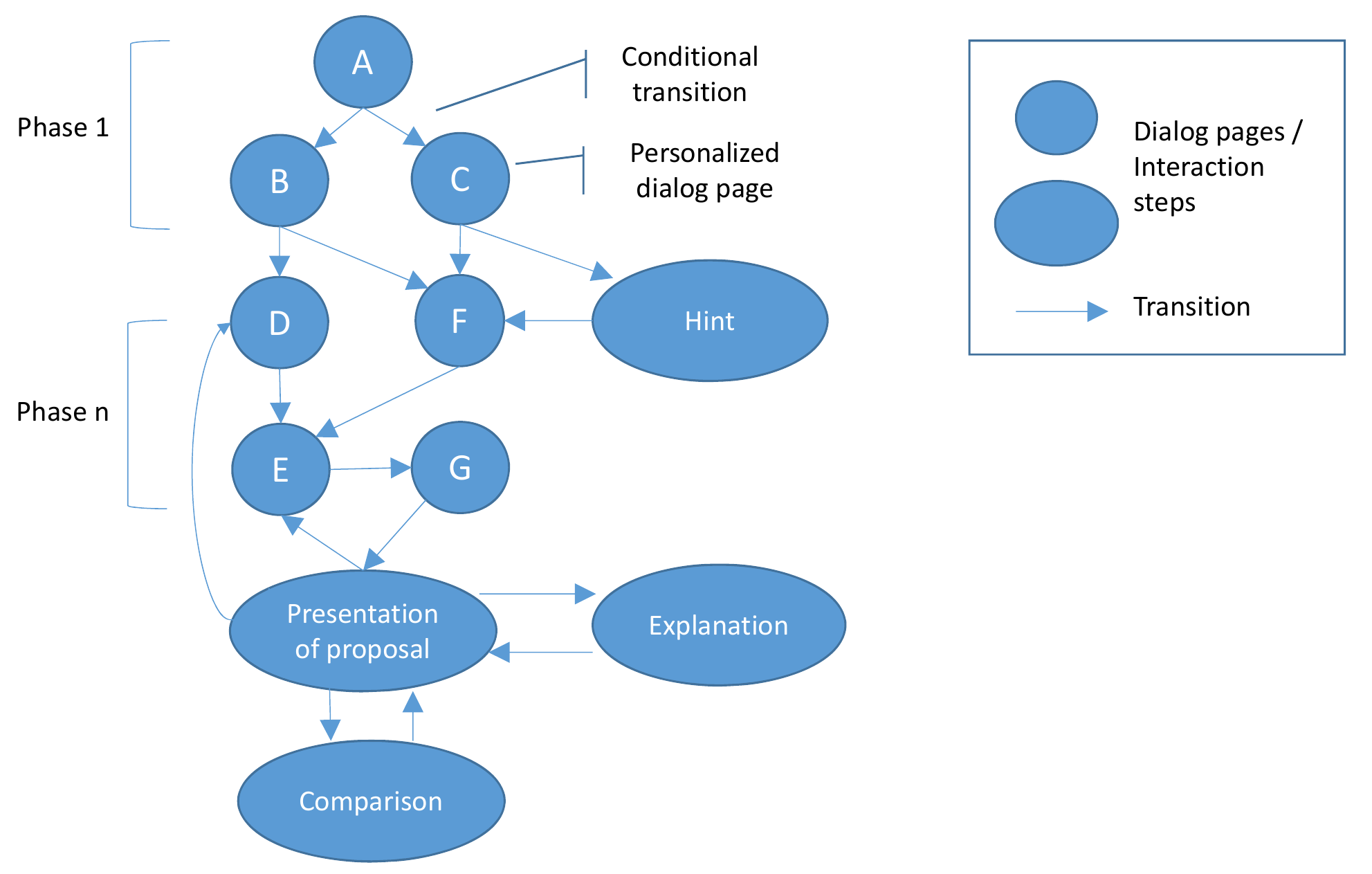}
  \caption{Pre-defined dialogue states in the Advisor Suite system (adapted from \citep{Jannach:2007:RDK:2011243.2011249}).}
  \label{fig:state-machine}
\end{figure}

Figure \ref{fig:state-machine} shows a schematic overview of such a dialogue model. It consists of
\begin{enumerate*}[label=\textit{(\roman*)}]
  \item a number of dialogue steps 
  that to acquire the user's preferences through questions, and
  \item special dialogue states in which the system presents the results, provides explanations, or shows a comparison between different alternatives.
\end{enumerate*}
The possible transitions are defined at design time, but which path is taken during the dialogue is determined dynamically based on decision rules.
Another example for a work that is based on a pre-defined set of states and possible transitions is the interactive tourism recommender system proposed in \citep{Mahmood:2009:IRS:1557914.1557930}. In their case, the transitions at run-time are not determined based on manually engineered decision rules, but learned from the data using reinforcement learning techniques, where one goal is to minimize the number of required interaction steps.

Technically, there are different ways of explicitly representing such state machines. Some tools, as the one mentioned above, use visual representations, others rely on textual and declarative representations like ``dialogue grammars'' \citep{DBLP:conf/ewcbr/Bridge02} and case-frames \cite{Bobrow1977GUSAF}. Google's DialogFlow\footnote{\url{https://dialogflow.com/}}, as an example of a commercial service, uses a visual tool to model linear and non-linear conversation flows, where non-linear means that there are different execution paths, depending on the user's responses or contextual factors. Finally, in some cases, the possible states are simply hard-coded as part of the general program logic of the application.

In some works, and in particular in early critiquing-based ones which are based on forms and buttons \cite{DBLP:journals/expert/RicciN07, smyth2004compound, reilly2007comparison}, only a few generic dialogue states exist, which means that no complex flow has to be designed. After an initial preference elicitation stage, recommendations are presented, and the system offers a number of critiques that the user can apply until a recommendation is accepted or rejected. Dialogue state management is therefore in some ways relatively light-weight. The main task of the system in terms of dialogue management is to keep track of the user responses and, in case of dynamic critiquing, make inferences regarding the next critiques to offer.

Similarly, in some NLP-based conversational preference elicitation systems such as  \citep{christakopoulou2016towards,Zhang:2018:TCS:3269206.3271776}, there are mainly two phases: asking questions, in this case in an adaptive way, and presenting a recommendation list. 
In other NLP-based systems, the possible dialogue states are not modeled explicitly as such, but implicitly result from the implemented intents. For example, whether or not there is a dialogue state ``provide explanation'' depends on the question whether a corresponding intent was considered in the design phase of the system.

Finally, in the NLP-based end-to-end learning CRS proposed in \cite{li2018towards}, the dialogue states are in some ways also modeled implicitly, but in a different way. This system is based on a corpus of recorded human conversations (between crowdworkers) centered around movie recommendations. This corpus is used to train a complex neural model, which is then used to react to utterances by users. Looking at the conversation examples, these conversations, besides some chit-chat, mainly consist of interactions where one communication partner asks the other if she or he likes a certain movie. The sentiment of the answer of the movie seeker is then analyzed to make another recommendation, again mostly in the form of a question. The dialogue model is therefore relatively simple and encoded in the neural model. It seemingly does not support many other types of intents or information requests that do not contain movie names (e.g., \emph{``I would like to see a sci-fi movie''}).

\subsection{Background Knowledge}
Besides the discussed knowledge regarding the set of supported user intents or the possible dialogue states, CRS are based on additional types of knowledge and data. This knowledge for example includes information related to the items (e.g., attributes and ratings), corpora of logged natural language conversations for learning, and additional knowledge sources used for entity recognition.

\subsubsection{Item-related Information}
Like any recommender, also a conversational system has to have access to a database with information about the recommendable items. Such a database can contain item ratings,  meta-data that can be presented to the user (e.g., the genre of a movie or the director), community-provided tags, or extracted keyphrases. These item attributes can furthermore serve as a basis for other computational tasks, e.g., to compute the personalized recommendations, to generate explanations, or to determine which questions can be asked to the user.

In the examined papers, we found that researchers used a number of different databases. Some works are based on typical rating datasets, e.g., from MovieLens or Netflix, whereas other researchers created their own datasets or relied on preexisting datasets from different domains. In Table \ref{tab:item-datasets}, we provide examples of datasets containing item-related information. It can be observed that it is not uncommon, e.g., in critiquing-based applications, that researchers solely rely on datasets which they created or collected for the purpose of their studies, i.e., there is limited reuse of datasets by other researchers.   One main underlying reason is that in most papers we analyzed, researchers did not publicly share their datasets.

\begin{table} [h!t]
  \centering
    \caption{Examples of datasets containing item-related information.}
    \label{tab:item-datasets}
\small
\begin{tabular}{p{3cm}p{10cm}}
\hline
\textbf{Domain} & \textbf{Description}\\
\hline
Movies & Traditional movie rating databases from MovieLens, EachMovie, Netflix, used for example in \cite{li2018towards,Zhao:2013:ICF:2541154.2505690,Zhao:2013:ICF:2541154.2505690}. \\ \hline
Electronics &  A product database with more than 600 distinct products was collected from various retailers \cite{Grasch:2013:RTC:2507157.2507161}. \\ \cline{2-2}
      & A smartphone database consisting of 1721 products with multiple features \cite{Contreras2018}. \\ \cline{2-2}
      & An Amazon electronics review dataset
      containing millions of products, user reviews and product meta-data \cite{Zhang:2018:TCS:3269206.3271776}. \\ \cline{2-2} 
    & A dataset consisting of 120 personal computers, each with 8 features \cite{smyth2004compound}. \\ \hline
Travel & More than 100 sightseeing spots in Japan with 25 different features \cite{Ikemoto2019}. \\ \cline{2-2}
     & A database of restaurants in the San Francisco area covering 1,900 items with multiple features like cuisine, ratings, price, location, or parking  \cite{Thompson:2004:PSC:1622467.1622479}. \\ \cline{2-2}
     & Search logs and reviews of 3,549 users of a restaurant review provider, focusing on locations in Cambridge \cite{christakopoulou2016towards}.
     \\      \cline{2-2}
    & A travel destinations dataset, crawled from online platforms containing 5,723,169 venues in 180 cities around the globe \cite{Dietz2019}. \\ \cline{2-2}
    & A restaurants dataset crawled for Dublin city, which consists of 632 restaurants with 28 different features \cite{mccarthy2010experience}. \\ \hline
Food Recipes &  A food recipe dataset containing dishes and their ingredients \cite{zeng2018eliciting}. \\ \hline
E-commerce & A product database of 11M products and logged data from the search engine of an e-commerce website was collected. The logged data consists of 3,146,063 unique questions \cite{DBLP:conf/aaai/YanDCZZL17}.\\ \hline

Music &  A music dataset crawled from multiple online sources, containing 2,778 songs with 206k explanatory statements and 22 user tags \cite{zhao2019personalized}. \\ \hline
\end{tabular}
\normalsize
\end{table}

\subsubsection{Dialogue Corpora Created to Build CRS}
\label{subsec:dialogue-corpora}
NLP-based dialogue systems are usually based on training data that consist of recorded and often annotated conversations between humans (interaction histories). A number of initiatives were therefore devoted to create such datasets that can be used to build CRS. Other researchers, in contrast, rely on dialogue datasets that were created or collected for other purposes. Generally, these corpora can be obtained with the help of crowdworkers \cite{Sun:2018:CRS:3209978.3210002, li2018towards, kang2019recommendation}, by annotating interviews \cite{cai2019department,DECAROLIS201787,Pecune:HAI2019}, or by logging interactions with a chatbot like in \cite{DBLP:journals/corr/JoshiMF17}. Table \ref{tab:dialogue-datasets} shows examples of such datasets used in recent research.

Note that in some cases when building a CRS, these dialogue corpora are combined with other knowledge bases \cite{li2018towards, zeng2018eliciting}. In \cite{li2018towards}, for example, both a dialogue corpus and MovieLens data are used for the purposes of sentiment analysis and rating prediction. Such a combination of datasets can be necessary when there is not enough relevant information in the dialogues.

\begin{table} [ht!]
  \centering
    \caption{Examples of dialogue corpora created or used to build CRS.}
    \label{tab:dialogue-datasets}
\small
\begin{tabular}{p{2cm}p{1.5cm}p{9.3cm}}
\hline
 \textbf{Domain} & \textbf{Name} & \textbf{Description} \\
\hline
Movies          & ReDial & Crowdworkers from Amazon Mechanical Turk (AMT) were used to collect over 10,000 dialogues centered around the theme of providing movie recommendations \cite{li2018towards}. A paired mechanism was used where one person acts as a \textit{recommendation seeker} and the other as a \textit{recommender}.
\\ \cline{2-3}
                & CCPE-M & A Wizard-of-Oz (WoZ) approach is taken to elicit movies preferences from crowdworkers within natural conversations. The dataset consists of over 500 dialogues that contain over 10,000 preference statements \cite{radlinski2019coached}. \\ \cline{2-3}
                & GoRecDial & This dataset consists of 9,125 dialogue interactions and 81,260 conversation turns collected through pairs of human workers; here also one plays the role of a movie seeker and the other as a recommender \cite{kang2019recommendation}.  \\ \cline{2-3}
                & bAbI & In \cite{DBLP:conf/recsys/NarducciBIGLS18}, the authors used a general movie dialogue dataset provided by Facebook Research \cite{DBLP:journals/corr/DodgeGZBCMSW15} to build a CRS. The dataset contains task-based conversations in a question-answering style. It consists of 6,733 and 6,667 dialogue conversations for training and testing respectively. \\
                \hline
Restaurants and Travel    & CRM & An initial dataset containing 385 dialogues is collected using a pre-defined dialogue template through AMT \cite{Sun:2018:CRS:3209978.3210002}. Using this dataset, a larger synthetic dataset of 875,721 simulated dialogues is created. \\ \cline{2-3}
                & ParlAI &  A goal-oriented, extended version of the bAbI dataset that was collected using a bot and users. It consists of three datasets (training, development and testing), each comprising 6,000 dialogues. \cite{DBLP:journals/corr/JoshiMF17}. \\ \cline{2-3}
            & MultiWOZ & A large human-human dialogue corpus, which covers 7 domains and consists of 8,438 multi-turn dialogues around the themes of travel \& planning recommendation \cite{wu2019transferable}. \\ \hline
Fashion         & MMD & A dataset consisting of 150,000 conversations between shoppers and a large number of expert sales agents is collected. 9 dialogue states were identified in the resulting dataset \cite{Saha:1704.00200}. \\ \hline
Multi-domain  &  OpenDialKG & Chat conversation between humans, consisting of 15,000 dialogues and 91,000 conversation turns on movies, books, sports, and music \cite{moon2019opendialkg}.
 \\ \hline
\end{tabular}
\normalsize
\end{table}

\subsubsection{Logged Interaction Histories}
Building an effective CRS requires to understand the conversational needs of the users, e.g., how they prefer to provide their preferences, which intents they might have, and so on. One way to better understand these needs is to log and analyze interactions between users and a prototypical system. These logs then serve as a basis for further research. Differently from the dialogue corpora discussed above, these datasets were often not primarily created to build a CRS, but to better understand the interaction behavior of users. In \cite{warnestaal2007interview, warnestaal2005user}, for example, the interactions of the user with a specific NLP-based CRS were analyzed regarding dialogue quality and dialogue strategies. In \cite{cai2019department, DECAROLIS201787}, user studies were conducted prior to developing the recommender system to understand and classify possible feedback types by users. In some approaches like \cite{DECAROLIS201787, qiu2017alime} researchers annotated and labeled such datasets for the purpose of model training and system initialization. However, such logged histories are---except for \cite{qiu2017alime}---typically much smaller in size than the dialogue corpora discussed above, mostly because they were collected during studies with a limited number of participants. Examples of datasets obtained by logging system interactions and user studies are shown in Table \ref{tab:logged-interactions}.

\begin{table} [h!t]
  \centering
    \caption{Examples of datasets obtained from logged system interactions and user studies.}
    \label{tab:logged-interactions}
\small
\begin{tabular}{p{2cm}p{10cm}}
\hline
\textbf{Domain} & \textbf{Description}\\
\hline
Movies & A dialogue dataset involving 347 users was collected in \cite{kang2017understanding} during the experimental evaluation of a recommender system. \\
    \cline{2-2}
    & A subset of the ReDial dataset was analyzed and annotated in \cite{cai2019department} to classify the user feedback types in 200 dialogues at the utterance level.\\ \cline{2-2}
    & A dialogue corpus was collected in \cite{warnestaal2005user} for the purpose of dialogue quality analysis
    consisting of 226 complete dialogue turns with 20 users. \\ \cline{2-2}
   & A user study
   was conducted in \cite{warnestaal2007interview}, where a \textit{movie seeker} and a \textit{human recommender} converse with each other. The dialogue corpus consists of 2,684 utterances and 24 complete dialogues.\\ \hline
Travel & A dataset containing preferences for hotel, flight, car rental searches was collected in \cite{Argal2018} involving 200 users of a content-based recommender system that supports multiple tasks (i.e., hotel, car, flight booking) in the same dialogue.\\ \hline
Fashion & A user study was conducted using a virtual shopping system. 
A non-verbal feedback (e.g., gestures, facial expressions, voices) dataset involving 345 subjects was collected and then annotated for model training \cite{DECAROLIS201787}. \\ \hline
E-commerce & A dataset containing conversation logs of users with a chatbot of an online customer service center (Alibaba.com) was collected in \cite{qiu2017alime}. It consists of over 91,000 Q\&A pairs as a knowledge base used for the information retrieval task. \\ \hline
\end{tabular}
\normalsize
\end{table}

\subsubsection{Lexicons and World Knowledge}
Researchers often use additional knowledge bases to support the entity recognition process in NLP-based systems.
In \cite{DBLP:conf/aaai/LeeMRAJ18,liu2010dialogue}, for instance, information was harvested from online sources such as Wikipedia or Wikitravel to develop dictionaries for the purpose of entity-keyword mapping.
Similarly, the WordNet corpus was used in \cite{DBLP:conf/aaai/LeeMRAJ18} to determine the semantic distance of an identified keyword in a conversation with predefined entities.
More examples for the use of lexicons and world knowledge are shown in Table \ref{tab:world-knowledge-datasets}.


\begin{table} [h!t]
  \centering
    \caption{Examples of the use of lexicons and world knowledge.}
    \label{tab:world-knowledge-datasets}
\small
\begin{tabular}{lp{11cm}}
\hline
\textbf{Source Name} & \textbf{Description} \\
\hline
Wikipedia            & A dataset crawled from online sources (Wikipedia and Wikitravel) for the purpose of entity recognition in the travel domain \cite{DBLP:conf/aaai/LeeMRAJ18}.\\
WordNet              & WordNet
is used in order to compute the semantic distance between entities and keywords mentioned in the conversation \cite{DBLP:conf/aaai/LeeMRAJ18,liu2010dialogue}. \\
Wikiquote            & A quote dataset crawled from two online sources, wikiquote.com and the Oxford Concise Dictionary of Proverbs \cite{lee2016quote}.
\\
Citysearch           & In \cite{liu2010dialogue}, a dataset of 137,000 users reviews on 24,000 restaurants was harvested from two online sources (citysearch.com and menupages.com to generate a dictionary of mappings between semantic representations of cuisines and dialogue concepts.                                                                                   \\ \hline
\end{tabular}
\normalsize
\end{table}

\subsection{Discussion}
Our discussions show that CRS can be knowledge-intensive or data-intensive systems. Differently from the traditional recommendation problem formulation, where the goal is to make relevance predictions for unseen items, CRS often require much more background information than just a user-item rating matrix, in particular in the context of dialogue management.

\paragraph{Pre-defined Knowledge vs.~Learning Approaches.} In CRS approaches that use forms and buttons as the only interaction mechanism, the interaction flow is typically pre-defined in the form of 
the possible dialogue states, the set of supported user intents, and the user profile attributes to acquire. NLP-based systems, in contrast, are usually more dynamic in terms of the dialogue flow, and they rely on additional knowledge sources like dialogue corpora and answer templates as well as lexicon and word knowledge bases. Nonetheless, these systems typically require the manual definition of additional background knowledge, e.g., with respect to the supported user intents. 

Pure ``end-to-end'' learning only from recorded dialogues seems challenging. In most existing approaches the set of supported interaction patterns is implicitly or explicitly predefined, e.g., in the form of ``user provides preferences, systems recommends''. To a certain extent, also the collection of human-to-human dialogues can be designed to support possible system responses like in \cite{li2018towards}, where the crowdworkers were given specific instructions regarding the expected dialogues. As a result, the range of supported dialogue utterances can be relatively narrow. The system presented in \cite{li2018towards}, for example, cannot handle a query like \emph{``good sci-fi movie please''}.

\paragraph{Intent Engineering and Dialogue States.} In case a richer dialogue and additional functionalities are desirable, the definition of the supported user intents usually is a central and often manual task during CRS development.
Compared to general-purpose dialogue systems and home assistants, however, the set of user intents that will be supported is often relatively small. We have identified some common intent categories in Section \ref{subsec:supported-user-intents}. Depending on the domain, also very specific intents can be supported, e.g., asking for a style tip in a fashion recommender system \citep{nie2019multimodal}. Furthermore, yet another set of possible user intents has to be supported in CRS that are designed for group decision scenarios. Typical user intents can, for example, relate to the invitation of collaborators \citep{nguyen2017chat} or to a request for a group recommendation. Furthermore, there might be user utterances that relate to the resolution of preference conflicts and voting among group members \citep{mccarthy2006group, Alvarez2016,nguyen2017chat}.

Generally, the set of user intents that the system supports determines how rich and varied the resulting conversations can be. Not being able to appropriately react to user utterances can be highly detrimental to the quality perception of the system. For example, being able to explain the recommendations that the system makes is often considered as a key feature to make decision-making easier or to increase  user trust in a recommender system. A user of an NLP-based system might therefore be easily disappointed by the conversation if the system fails to recognize and respond to a request for an explanation.

A key challenge therefore is to anticipate or learn over time which intents the users might have. Depending on the application and used technology, the design and implementation of an intent database (e.g., using Google's DialogFlow engine) can lead to substantial manual efforts
and require the involvements of professional writers to achieve a certain naturalness and richness of the conversation. 
At the same time, the rule-based modeling approach (``if-this-then-that'') as implemented by major solution providers can easily lead to large knowledge bases that are difficult to maintain, leading to a need for alternative modeling approaches \cite{Telang2018}.

\section{Computational Tasks}
\label{sec:computational-tasks}
Having discussed possible user intents in recommendation dialogues, we will now review common computational tasks and technical approaches for CRS. We distinguish between
\begin{enumerate*}[label=\textit{(\roman*)}]
  \item main tasks, i.e., those related more directly to the recommendation process, e.g., compute recommendations or determine the next question to ask, and
  \item additional, supporting tasks.
\end{enumerate*}

\subsection{Main Tasks}
\begin{figure}[h!t]
\centering
\includegraphics[width=0.75\linewidth]{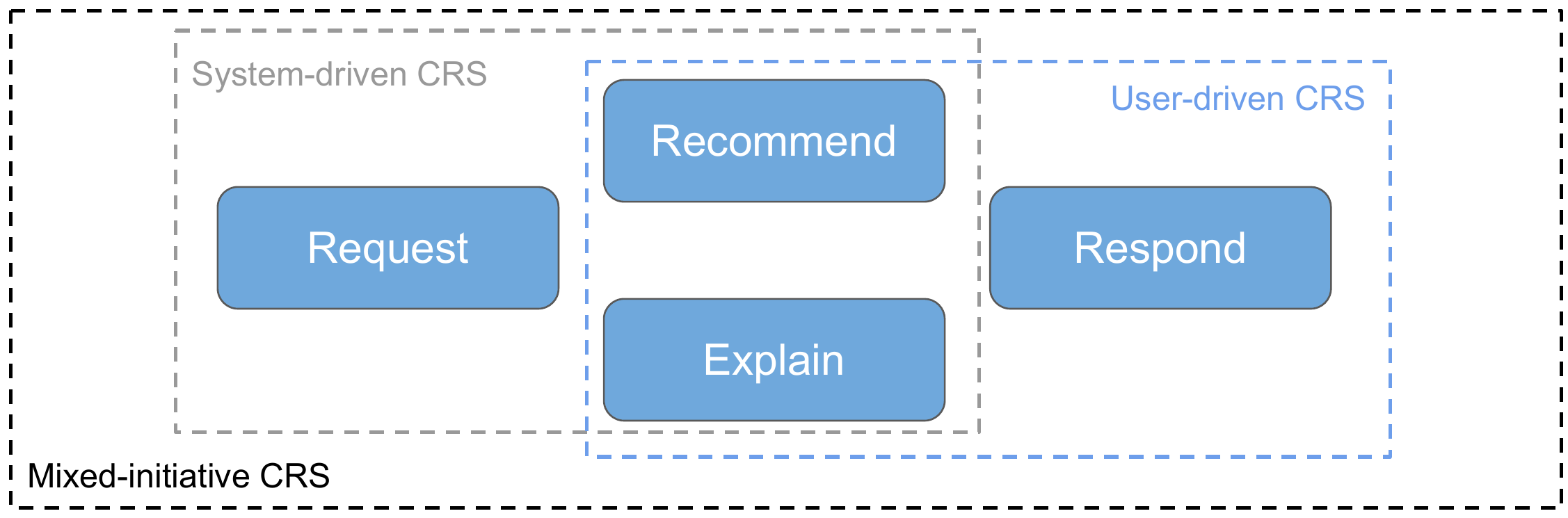}
\caption{Major actions taken by conversational recommender systems.}
\label{fig:recommending-action}
\end{figure}

Broadly speaking, CRS carry out four general types of tasks (or: system actions) during conversations \cite{Narducci2018aiai, chenpucritiquing2012}:  \emph{Request},  \emph{Recommend}, \emph{Explain}, and \emph{Respond}, see Figure \ref{fig:recommending-action}. However, not every CRS necessarily implements all of them. System-driven CRS (\textcolor{blue}{as described in Section 3.3})
usually drive the conversation by \textcolor{blue}{requesting user preferences on attributes and allowing users to give feedback on recommendations through multiple interaction cycles}. User-driven systems, in contrast, can take a more passive role, and mainly respond to conversational acts by the user. In mixed-initiative systems, e.g., those based on natural language interfaces, all types of actions can be found.

\subsubsection{Request}
A number of CRS follow a ``slot-filling'' conversation approach where the system seeks to acquire preference information about a pre-defined set of item attributes or facets. One main computational task in this context is to \emph{determine the next question to ask}, often with the goal to increase dialogue efficiency, i.e., to minimize the number of required interactions (see also Section \ref{sec:evaluation}). Various methods to determine the order of the facets were proposed in the literature \cite{Shimazu2002,Thompson:2004:PSC:1622467.1622479,Chakraborty2019RecommendenceAF,zeng2018eliciting}. In an early system \cite{Thompson:2004:PSC:1622467.1622479}, specific weights were used to rank the item attributes for which the user has not expressed preferences yet. Entropy-based methods also consider the potential effects on the remaining item space of each attribute. They aim to identify the next question (attribute) that helps mostly to narrow down the candidate (item) space \cite{DBLP:conf/eee/MirzadehRB05,Shimazu2002,Chakraborty2019RecommendenceAF, nica2018chatbot,Yin:2017:DID:3097983.3098148}, sometimes including feature popularity information \cite{DBLP:conf/eee/MirzadehRB05}.
Considerations like this are typically also the foundation of typical \emph{dynamic} and \emph{compound} critiquing systems \cite{mccarthy2004thinking,Reilly:ACBR2004,smyth2004compound,DBLP:conf/ah/ZhangP06,chen2007preference,DBLP:conf/recsys/ViappianiPF07,Pu:CHI2006}. In compound critiquing systems, in particular, the user is not asked about feedback for one single attribute, but for more than one within one interaction, e.g., \emph{``Different Manufacturer, Lower Processor Speed and Cheaper''}. Finally, in some systems, possible sequences of questions asked to the users are pre-defined in the form of state machines \citep{Jannach:2007:RDK:2011243.2011249,Jannach:2004:ASK:3000001.3000153}. At run-time, the dialogue path is then chosen based on the users' inputs in the ongoing session.

Instead of using heuristics for attribute selection and static dialogue state transition rules, a number of more recent systems rely on learning-based approaches, e.g., using reinforcement learning \cite{Sun:2018:CRS:3209978.3210002,tsumita2019dialogue,Mahmood:2009:IRS:1557914.1557930}.
In \cite{Sun:2018:CRS:3209978.3210002}, for example, the authors use a deep policy network to decide on the system action. Based on the current dialogue state, as modeled by a belief tracker, the system either makes a request for a pre-defined facet or generates a recommendation to be shown to the user. An alternative learning-based way to determine the question order was proposed in \cite{Christakopoulou:2018}. In their work, the authors design a recommender for YouTube that leverages past watching histories of the user community and a Recurrent Neural Network architecture to rank the questions (topics) that are shown to the user in a conversational step.

An alternative to asking users about attribute-based preferences is to ask them to give feedback on selected items. This can be done either by asking them to rate individual items (e.g., by like/dislike statements) or by asking them to express their preference for item pairs or entire sets of items \cite{DBLP:conf/chi/LoeppHZ14}. The computational task in this context is to determine the most informative item(s) to present to the user. Possible strategies include the selection of popular \textcolor{blue}{or diverse} items in the cold-start phase, items that are \textcolor{blue}{different} in terms of their past ratings or attributes, or itemsets that represent a balance of popularity and diversity \cite{Narducci2018aiai,DBLP:conf/iccbr/McGintyS03, DBLP:conf/iui/CareniniSP03,AM:IUI2002}.
However, not only item features might be relevant for the selection of the items. In \cite{DBLP:conf/iui/CareniniSP03}, the authors found that a user's willingness to give feedback on an item can depend on additional factors.
Specifically, they identified several situations in which the feedback probability may be higher, e.g., when the system's predicted rating deviates from the user's past experience of the item.
In more recent works, again learning-based approaches are more common. The authors of \cite{christakopoulou2016towards, Zhao:2013:ICF:2541154.2505690}, for example, employed bandit-based approaches
to either
\begin{enumerate*}[label=\textit{(\roman*)}]
  \item determine the next item to be shown for eliciting the user's absolute feedback (i.e., like or dislike), or
  \item to select a pair of items for obtaining the user's relative preference regarding these two items.
\end{enumerate*}

\subsubsection{Recommend}
The recommendation of items is the core task of any CRS. From a technical perspective, we can find collaborative, content-based, knowledge-based, and hybrid approaches in the literature. Differently from non-conversational systems, the majority of the analyzed CRS approaches mainly relies solely on short-term preference information. However, there are also approaches that additionally consider long-term preferences of a user, e.g., to speed up the elicitation process \cite{Sun:2018:CRS:3209978.3210002,Thompson:2004:PSC:1622467.1622479,warnestaal2005user,MapBasedRicci2010,Stanley2010,Yasser:WWW2016:HistoryGuided,nguyen2017chat}.

In the context of critiquing-based and knowledge-based systems, different strategies are applied to filter and rank the items. For the filtering task, often constraint-based techniques \cite{recsyshandbook2015constraints} are applied that remove items from the candidate set which do not (exactly) match the current user's preferences. The items that remain can then be sorted in different ways \cite{DBLP:journals/umuai/ZankerJ09}. In the system proposed in \cite{DBLP:conf/ah/ZhangP06}, for example, the user preference model is updated after a user critique by adjusting the weights of the attributes that are involved in the critique. Then, Multi-Attribute Utility Theory (MAUT) \cite{keeney1976decisions} was used to calculate the utility of each candidate item for generating top-K recommendations for the user. An alternative ranking approach was applied in \cite{Yasser:WWW2016:HistoryGuided}, where a history-guided critiquing system was proposed that aims to retrieve recommendation candidates from other users' critiquing sessions that are similar to the one of the current user. In \cite{Dietz2019}, a critiquing-based travel recommender system was implemented that computes recommendations based on the relevance of item attributes to user preferences based on the Euclidean Distance.

Some works consider both long-term and short-term preferences of users when making recommendations \cite{DBLP:journals/expert/RicciN07,Stanley2010,Yasser:WWW2016:HistoryGuided, Argal2018}. The \textsc{\small{Adaptive Place Advisor}} system \cite{Thompson:2004:PSC:1622467.1622479} represents an early example of combining short-term and long-term preferences. Here,  the user's current query is expanded by considering the probability distribution of the user's past preference for item attributes, based on her/his short-term constraints (within a conversation) and long-term constraints (over many conversations). This expanded query was then used to retrieve and rank the items for recommendation. {In \cite{Yasser:WWW2016:HistoryGuided}, the authors proposed to leverage the successful recommendation sessions in the previous conversations to improve the efficiency of the current session (i.e., to shorten its length).

More recent works rely on machine learning models and background datasets for the recommendation task. One common approach is to train a model on the traditional user-item interaction matrix, e.g., based on probabilistic matrix factorization \cite{christakopoulou2016towards}, and to then combine the user's current interactions with the trained user and item embeddings. In another approach \cite {Argal2018}, the authors rely on a content-based method based on item features and the user profile in the cold-start stage, and then switch to a Restricted Boltzmann Machine collaborative filtering method once a sufficient number of preference signals is available. In \cite{Zhang:2018:TCS:3269206.3271776}, a hybrid multi-memory network with attention mechanism was trained to find suitable recommendations based on item embeddings and the user's query embedding. Here, the item embedding was based on the item's textual description, and the user's query embedding encoded the user's initial request and the follow-up conversations during the interaction. A hybrid model was also proposed in \cite{Sun:2018:CRS:3209978.3210002}, which used Factorization Machines to combine the 
dialogue state---represented with an LSTM-based belief tracker for each item facet---user information, and item information to train the recommendation model. In the video recommender system presented in \cite{Christakopoulou:2018}, finally, an RNN-based model was built for making recommendations, based on the topics selected by the users and their watching history.

In some cases, application-specific techniques were applied for the recommendation task. In \cite{Yu:2019:VDA:3292500.3330991,Tong:MM2019:VisionLanguage}, for example, the CRS features a visual dialogue component, where users can give feedback based on the images, e.g., \emph{``I prefer blue color''}. To implement this functionality, the system proposed in \cite{Yu:2019:VDA:3292500.3330991} implemented a component that encoded item images and user feedback using a convolutional neural network, and then combined these encodings as an input to both a response encoder and a state tracker. Furthermore, various types of user behaviors (i.e., viewing, commenting, clicking) on the visually represented recommendation were considered in a bandit approach to balance exploration and exploitation.

\subsubsection{Explain}
The value of explanations in general recommender systems is widely recognized \cite{Herlocker:2000:ECF:358916.358995,Tintarev2011Expl,Explanations:UMUAI2017}. Explanations can increase the system's perceived transparency, user trust and satisfaction, and they can help users make faster and better decisions \cite{Gedikli2014367}. However, according to our survey, few papers so far have studied the explanation issue specific to CRS.
	
In the context of critiquing-based systems,
\cite{ExplainA} examined the trust-building nature of explanations. In this work, an ``organization-based'' explanation approach was evaluated, where the system showed multiple recommendation lists to the user, each of them labelled based on the critiquing-based selection criteria, e.g., \enquote{\emph{cheaper but heavier}}. A more recent interactive explanation approach for a mobile critiquing-based recommender was proposed in \cite{Lamche2014}, where the textual explanations to be shown to the user were determined based on the user's preferences and constructed from pre-defined templates.

Providing more information about a recommended item, e.g., in the form of pros and cons, is a typical approach when providing explanations. Generating such item descriptions in a user-tailored way in the context of CRS was proposed in \cite{Foster2010umuai} and \cite{WALKER2004811}. In such approaches, the users' feedback during the conversation can influence which attributes are mentioned in the item descriptions shown to the user in the recommendation phase. Furthermore, the user preferences can be considered to order the arguments and to help determine which adjectives and adverbs to use in the explanation \cite{Foster2010umuai}.

In \cite{Narducci2018aiai}, two kinds of explanations were implemented in a CRS for movies. One was simply based on the details of a given movie, whereas the other connects the given user preferences with item features through a graph-based approach to create a personalized explanation. 
Another graph-based approach following similar ideas was proposed in \cite{moon2019opendialkg}, where a knowledge-augmented dialogue system for open-ended conversations was discussed. In this approach, relevant entities and attributes in a dialogue context were retrieved by walking over a common fact knowledge graph, and the walk path was used to create explanations for a given recommendation. In \cite{Pecune:HAI2019}, finally, a human-centered approach was employed. By analyzing a human-human dialogue dataset, the authors identified different social strategies for explaining movie recommendations. They then accommodated the social explanation
in a conversational recommender system to improve the users' perception of the quality of the system.

However, for the main task of explaining, we found that little CRS-specific research exists so far, and only a smaller set of the proposed CRS in the literature support such a functionality.

\subsubsection{Respond}

This category of tasks is relevant in user-driven or mixed-initiative NLP-based CRS, where the user can actively ask questions to the system, actively make preference statements, or issue commands. The system's goal is to properly react to user utterances that do not fall in the above-mentioned categories ``Recommend'' and ``Explain''. Two main types of technical approaches can be adopted to respond to such user utterances. One approach---also commonly used in chatbots---is to map the utterances to pre-defined \emph{intents}, such as the ones mentioned in Table \ref{tab:user-intents}, e.g., Obtain Details or Restart. The system's answers to these pre-defined intents can be implemented in the system with the help of templates. In the literature, various user utterances are mentioned to which a CRS should be able to respond appropriately. Examples include preference refinements, e.g., \enquote{\emph{I like Pulp Fiction, but not Quentin Tarantino}} \cite{Narducci2018aiai}, requests for more information about an item, or a request for the system's judgement regarding a certain item, e.g., \enquote{\emph{How about Huawei P9 ?}} \cite{DBLP:conf/aaai/YanDCZZL17}.

An alternative technical approach is to select or generate the system's responses by automatically training a machine learning model from dialogue corpora and other knowledge sources like in ``end-to-end'' learning systems, e.g., \cite{li2018towards,chen-etal-2019-towards,DBLP:journals/corr/JoshiMF17,kang2019recommendation}.
In an open-domain dialogue system described in \cite{qiu2017alime}, for example, an information retrieval model was used to retrieve an initial set of candidate answers from a Q\&A knowledge base (an online customer service chat log). Then, an attentive sequence-to-sequence model was used to rank the candidate answers in order to determine answers with scores that are higher than a pre-defined threshold. If no existing answer was considered suitable, a sequence-to-sequence based model was used to generate the system's response.

Another example for such an approach is described in \cite{nie2019multimodal}. In this multi-modal recommender system, one RNN model was used to generate general responses such as greetings or chit-chat, and a knowledge-aware RNN model was trained to answer more specific questions. For instance, when the user asks for style-tip: \enquote{\emph{Will T-shirt complement any of these sandals?}}, the system may respond with \enquote{\emph{Yes, T-shirt will go well with these sandals}} \cite{nie2019multimodal}.

Finally, some approaches were proposed in the literature to deal with very specific dialogue situations. One example of such a situation is a \emph{conversational breakdown}, where the system is unable to understand the user's input \cite{Myers:CHI2018}. Possible repair strategies, such as politeness and apology strategies, were examined in the area of human-robot interaction to mitigate
the negative impact of such a breakdown \cite{LEE:HRI2010,Srinivasan:CHI2016, Lee:HRI2011}. Various repair strategies based on communication theory, e.g., repeating or asking for clarifications, or strategies from explainable machine learning, e.g., explaining which parts of the conversation were not understood, can in principle be applied \cite{ashktorab2019resilient}. %
\subsection{Supporting Tasks}
Depending on the system's functionality, a number of additional and supporting computational tasks may be relevant in a CRS.

\paragraph{Natural Language Understanding}
In NLP-based CRS, it is essential that the system understands the users' intents behind their utterances, as this is the basis for the selection of an appropriate system action \cite{rafailidis2018technological}. Two main tasks in this context are \emph{intent detection} and \emph{named entity recognition}, and typical CRS architectures have corresponding components for this task.
In principle, intent detection can be seen as a classification task (dialogue act classification), where user utterances are assigned to one or multiple intent categories \cite{Stolcke:CL2000}. Named entity recognition aims at the identification of entities in a given utterance into pre-defined categories such as product names, product attributes, and attribute values \cite{DBLP:conf/aaai/YanDCZZL17}.
	
Although intent detection and named entity recognition have been extensively studied in general dialogue systems \cite{Stolcke:CL2000}, there are few studies specific to CRS according to our survey, possibly due to the lack of a well-established taxonomy and large-scale annotated recommendation dialogue data. In an early approach \cite{Thompson:2004:PSC:1622467.1622479}, manually-defined \emph{recognition grammars} were used to map user utterances to pre-defined dialogue situations, which is comparable to using pre-defined intents as described above in the context of the \emph{Respond} task.
An example for a more recent approach can be found in \cite{DBLP:conf/aaai/YanDCZZL17}. Here, a natural language understanding component for intent detection, product category detection, and product attribute extraction was implemented in a dialogue system for online shopping. For instance, from the utterance \enquote{\emph{recommend me a Huawei phone with 5.2 inch screen}} the system should derive the intent \emph{recommendation}, the product category \emph{cellphone}, as well as the brand and the display size. To solve these tasks, the authors first collected product-related questions from queries posted on a community site, and then extracted intent phrases (e.g., \enquote{\emph{want to by}} and \enquote{\emph{how about}}) by using two phrase-based algorithms. A multi-class classifier was trained for intent detection of new user questions. As for product category detection, the authors employed a CNN-based approach that took the detected intent into account to identify the category of a mentioned product in a given utterance.

Neural networks were used also in other recent intent and entity recognition approaches \cite{tsumita2019dialogue,nie2019multimodal}. For example, a Multilayer Perceptron (MLP) was used to predict the probability distribution on a set of pre-defined intent categories in \cite{nie2019multimodal}. A sequence-to-sequence model was used in \cite{Yin:2017:DID:3097983.3098148} to reframe the user's query (e.g., \enquote{\emph{How to protect my iphone screen}}) into keywords (e.g., \enquote{\emph{iphone screen protector}}) that are then used in the recommendation process to identify candidate items.

Another supporting task in some applications is \emph{sentiment analysis}, see, e.g., \cite{iovine2020conversational,li2018towards,DBLP:conf/recsys/NarducciBIGLS18,nie2019multimodal,zhao2019personalized}.
One typical goal in the context of CRS is to understand a user's opinion about a certain item. For example, whenever an item---e.g., a movie---is mentioned in an utterance, the sentiment of the sentence can be used to approximate the user's feelings about the item. This sentiment can then be considered as an item rating, which can subsequently be used for recommending other items using established recommendation techniques.

\paragraph{Specific Recommendation Functionality}
Depending on the technical approach to generate recommendations, specific computational subtasks may be helpful or required to support the recommendation process. We will give examples from the context of critiquing-based approaches here. In \cite{DBLP:conf/ictai/TrabelsiWBR10, Blanco:Recsys2013}, for example, the goal is to make ``query suggestions'' to users, where the term ``query'' is equivalent to a critique or constraint on the item features. In the mentioned approaches, the query suggestions (or modifications) are based on an extended analysis of the satisfiability of a query (i.e., will the suggested query lead to any results) or dominance relations between possible query suggestions. Generally, such query suggestions can be particularly helpful for users who have difficulties expressing their preferences. In the field of information retrieval, many approaches to query suggestion were proposed to assist users in expressing their information needs, see, e.g., \cite{Sordoni:CIKM2015} for a more recent example. Limited work however exists so far to apply such ideas to the context of CRS.

A related problem in constraint-based CRS is that in some cases the user's expressed preferences lead to the situation that either too many items remain for recommendation or that no item is left. Different approaches were proposed in the literature for query \emph{relaxation}. In \cite{Thompson:2004:PSC:1622467.1622479}, for example, a relatively simple strategy was adopted to remove some constraints. More elaborated strategies were proposed in \cite{DBLP:conf/ewcbr/McSherry04a,Jannach2006} and \cite{RicciQueryManagement2003}. In this latter work \cite{RicciQueryManagement2003}, the authors also introduce the concept of ``query tightening''. Here, the idea is to add more constraints on item attributes in case the number of relevant items returned by the system would lead to a choice overload problem. Generally, like for the query suggestion approaches described above, similar query revision approaches (relaxation and tightening) were not explored to a large extent in the context of NLP-based CRS. An exception is the concept for a chatbot presented in \cite{nica2018chatbot}, where the system tries to first identify the cause of the unsuccessful query and then asks the user to remove some preferences and to rank the item features by importance. Finally, instead of returning an empty result and asking the user to revise the preferences, approaches exist that automatically relax constraints and inform the user, e.g., that \enquote{\textit{There are 10 cameras less than 300 euro but their resolution is between 1 and 4 mega-pixels}} \cite{mccarthy2004thinking,Jannach:2004:ASK:3000001.3000153}.

\subsection{Discussion} %
Our analysis shows that a wide range of technical approaches are used in the literature to support the main computational tasks of a CRS. For the problem of computing recommendations, for example, all sorts of approaches---collaborative, content-based, hybrid---can be used within CRS. However, for the main task of explaining, we found that little CRS-specific research exists so far, and only a smaller set of the proposed CRS in the literature support such a functionality.

Another observation is that dialogue management is often sketched as a conceptual architectural component, but it is then implemented either in a rather static way with pre-defined transitions, e.g, see \cite{Mahmood:2009:IRS:1557914.1557930, Zhang:2018:TCS:3269206.3271776}, or done implicitly during the intent recognition and mapping phase or determined by the choices of the preference acquisition strategy, e.g., slot-filling \cite{wu2019transferable}. In some cases, the possible dialogue states are furthermore quite limited, e.g., the system can either decide to ask questions or to provide a recommendation \cite{Sun:2018:CRS:3209978.3210002}. Technically, in a few cases intent recognition and dialogue flow management are based on commercial tools, e.g, see \cite{ Dalton:2018:VGC:3209978.3210168, Arteag2019}.

In general, with the growing spread of chatbot applications, several commercial companies such as Google, Microsoft, Facebook and  IBM, have released frameworks or public APIs that implement some of the mentioned computational tasks and allow developers to create their own chatbots.
These tools include Google's \emph{DialogFlow} system, Facebook's \emph{Wit.ai}, and IBM \emph{Watson Assistant}  \footnote{\url{https://dialogflow.com}, \url{https://wit.ai/}, \url{https://www.ibm.com/cloud/watson-assistant/}} and provide functionalities such as speech recognition, voice control, the identification of pre-defined intents from natural language utterances, dialogue flow management, response generation to specific intents, and the deployment of applications to commercial platforms. Examples of research works that used these services include \cite{Jin:2019:MEC:3357384.3357923,FuzzyBased2018,Dalton:2018:VGC:3209978.3210168,Arteag2019,SinghcinICD102019,kang2017understanding, Angara2017}. Frameworks for the development of conversational systems are also provided by Microsoft through its \emph{Bot Framework} and by Amazon for its Alexa assistant and smart speakers. A CRS for the travel domain that uses Amazon Echo smart speakers was, for example, presented in \cite{Argal2018}. In general, however, these frameworks and services usually do not implement functionality that is specific to recommendation problems, but are designed to build general-purpose conversational systems.

Besides companies, also some researchers release their NLP-based CRS to the public. Examples include the VoteGoat movie recommender \cite{Dalton:2018:VGC:3209978.3210168} and the ConveRSE framework for chatbot development \cite{Narducci2018aiai,iovine2020conversational}.

\section{Evaluation of Conversational Recommenders}
\label{sec:evaluation}
In general, recommender systems can be evaluated along various dimensions, using different methodological approaches \cite{Shani2011}.
First, when a system is evaluated in its context of use, i.e., when it is deployed, we are usually interested mostly in specific \emph{key performance indicators} (KPIs) that measure---through A/B testing---if the system is achieving what it was designed for, e.g., increased sales numbers or user engagement \cite{jannachjugovactmis2019}.
Secondly, user studies (lab experiments) typically investigate questions related to the \emph{perceived quality} of a system. Common quality dimensions are the suitability of the recommendations, the perceived transparency of the process or the ease-of-use, see also \cite{Pu:2011:UEF:2043932.2043962}.
Offline experiments, finally, do not involve users in the evaluation, but assess the quality based on \emph{objective metrics}, e.g., the accuracy of predicting heldout ratings in a test set, by measuring the diversity of recommendations, or by computing running times.

The same set of quality dimensions and research methods can also be applied for CRS. However, when comparing algorithm-oriented research and research on conversational systems, we find that the main focus of the evaluations is often a different one. Since CRS are highly interactive systems, questions related to human-computer interaction aspects are more often investigated for these systems. Furthermore, regarding the measurement approaches, CRS evaluations not only focus on task fulfillment, i.e., if a recommendation was suitable or finally accepted, but also on questions related to the efficiency or quality of the conversation itself.

\subsection{Overview of Quality Dimensions, Measurements, and Methods}
Through our literature review, we identified the following main categories of quality dimensions investigated in CRS:
\begin{enumerate}
\item \emph{Effectiveness of Task Support}: This category refers to the ability of the CRS to support its main task, e.g., to help the users make a decision or find an item of interest.
  \item \emph{Efficiency of Task Support}: In many cases, researchers are also interested to understand how quickly a user finds an item of interest or makes a decision.
  \item \emph{Quality of the Conversation and Usability Aspects}: Analyses in this category focus on the quality of the conversation itself and on the usability (ease-of-use) of the CRS as a whole.
  \item \emph{Effectiveness of Subtask}: A number of studies investigated in our survey focus on certain subtasks like intent or entity recognition.
\end{enumerate}

In each of these dimensions, a number of different measurements are considered in the literature. Task \emph{effectiveness}, for example, can be both measured objectively (through accuracy measures, acceptance or rejection rates) or subjectively (through surveys related to choice satisfaction or perceived recommendation quality). Task \emph{efficiency} is very often measured objectively through the number of required interaction steps and shorter dialogues are usually considered favorable. The \emph{quality of the conversation} is most often analyzed in terms of subjective assessments, e.g., with respect to fluency, understandability, or the quality of the responses. Finally, specific measurements for subtasks include intent recognition rates or the accuracy of the state recognition process.

From a methodological perspective, we found works that entirely relied on offline experiments, works that relied exclusively on user studies, and studies that combined both offline experiments with user studies. Reports on fielded systems and A/B tests are rare. Examples of such works that discuss deployed systems include \citep{qiu2017alime,DBLP:conf/aaai/YanDCZZL17,Christakopoulou:2018,JannachENTER2007,Jannach:2004:ASK:3000001.3000153,BotWheels2017,Chakraborty2019RecommendenceAF,nica2018chatbot}. However, the level of detail provided for these tests is often limited, partially informal, or only considers certain aspects like processing times. Finally, we also found works without any evaluation or where the evaluation was mostly qualitative or anecdotal \cite{DBLP:conf/aaai/LeeMRAJ18,widyantoro2014framework,Argal2018}.

In the experimental evaluations, all sorts of materials---in particular prototype applications---and datasets were used. As discussed in Section \ref{sec:knowledge-behind}, at least an item database is needed. Depending on the technical approach, also additional types of knowledge and data are used, such as logged conversations between humans, explicit dialogue-related knowledge such as supported intents etc.

In Figure \ref{fig:evaluation-overview}, we provide an overview of the most common evaluation dimensions and evaluation approaches, and give examples for typical measurements and datasets. In the following sections, we will discuss some of the more typical evaluation approaches in more detail.

\begin{figure}[h!t]
\centering
\includegraphics[width=0.99\linewidth]{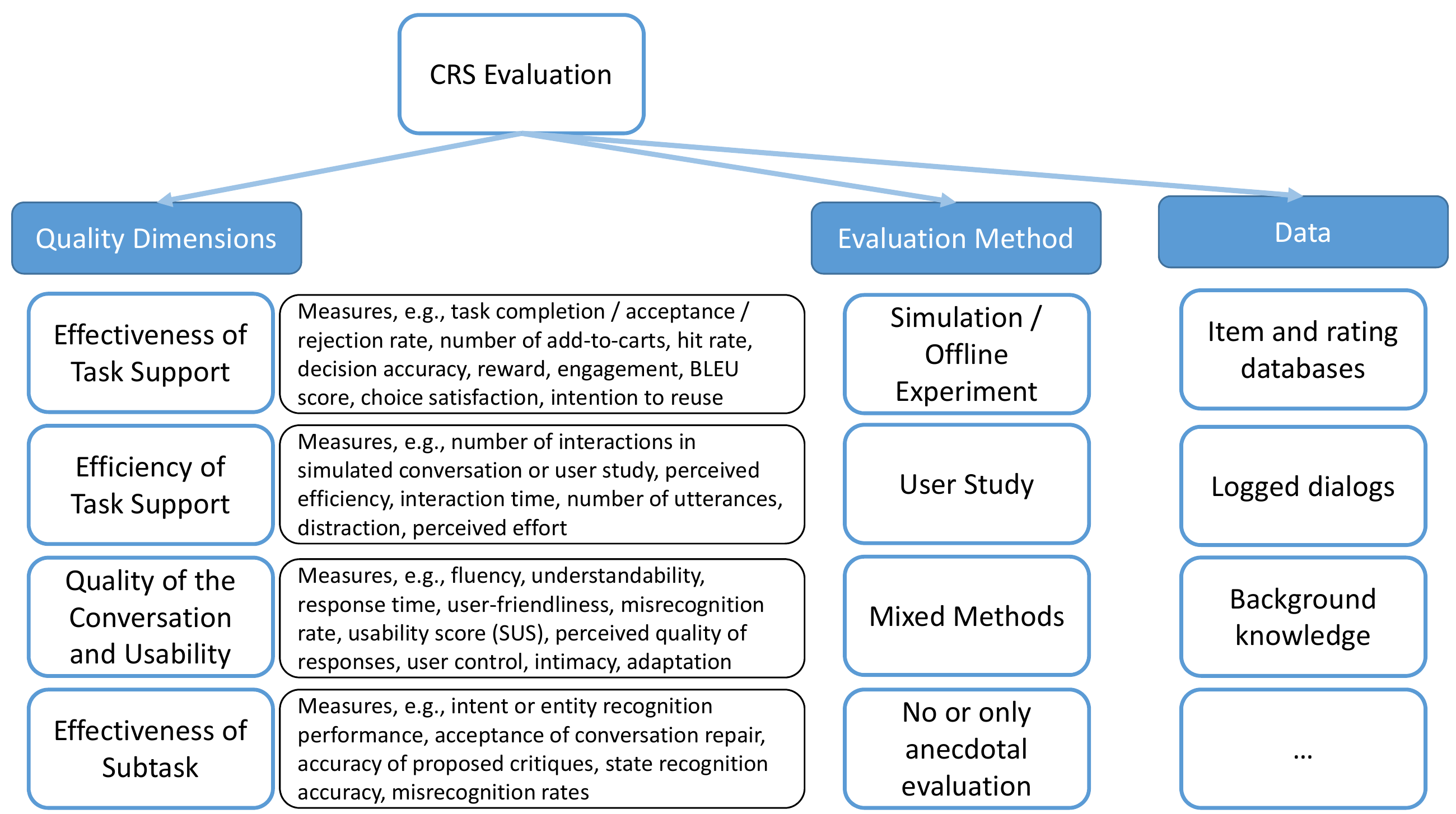}
\caption{Overview of evaluation dimensions and experiment designs.}
\label{fig:evaluation-overview}
\end{figure}

\subsection{Review of Evaluation Approaches}
\subsubsection{Effectiveness of Task Support}
In traditional recommender systems, the most common evaluation approach is to determine---through offline evaluations---how accurate an algorithm is at predicting some known, but withheld user preferences. The underlying assumption is that systems with higher accuracy are more effective, e.g., in helping users find what they need.
Objective accuracy measures such as the RMSE or the Hit Rate are sometimes also used to evaluate CRS. However, there are typically no long-term preferences available for conversational systems and the system only learns about the user preferences in the ongoing usage session. Therefore, alternative evaluation protocols are typically applied that rely on simulated users or user studies. Furthermore, researchers sometimes use specific objective metrics besides accuracy, and they also frequently rely on subjective quality assessments from users \cite{chen-etal-2019-towards,li2018towards,Sun:2018:CRS:3209978.3210002}. The objective and subjective quality measures are discussed below in detail.

\paragraph{Objective Measures}
Accuracy measures like Average Precision, the Hit Rate or RMSE were for example used as part of the evaluations in \cite{christakopoulou2016towards,DECAROLIS201787,Narducci2018aiai} or \cite{tsumita2019dialogue}.
In \cite{christakopoulou2016towards}, a framework for interactive preference elicitation was proposed that learns which questions should be asked to users in the cold-start phase. To evaluate different strategies, the authors use real and simulated user profiles and report the average precision of the recommendations after each question-answering round.
Similarly, a user simulator was used for the evaluation of a dialogue-based facet-filling recommender system based on deep reinforcement learning and end-to-end memory networks in \cite{tsumita2019dialogue} and \cite{Sun:2018:CRS:3209978.3210002}. The simulator in \cite{tsumita2019dialogue} was based on real user utterances extracted from a dataset about restaurant reservations \cite{DBLP:journals/corr/JoshiMF17}.
The objective measures included the recommendation accuracy (median of ranking and success rate), as well as the proportion of the simulated users who accepted the recommendations. In \cite{Sun:2018:CRS:3209978.3210002}, the ``online'' experiments were based on a dataset collected through crowdworkers
and the objective measures included Average Reward of the reinforcement learning strategy and the Success Rate (conversion rate), i.e., the fraction of successful dialogues.

The authors of \cite{Narducci2018aiai} present a domain-independent CRS framework, and they use the Hit Rate to assess the effectiveness of different system components such as the recommendation algorithm or the intent recognizer. To make the measurements, they use the above-mentioned bAbI dataset as a ground truth, where each example contains the user preferences, the recommendation request and the recommended item. A similar evaluation approach based on ground truth information derived from different real-world dialogues and accuracy measures (RMSE, Recall, Hit Rate) was adopted in \cite{li2018towards,chen-etal-2019-towards,kang2019recommendation}. In such approaches, the system typically analyzes (positive) mentions of items (movies) in the ongoing natural language dialogue and use these preferences for the prediction task.

The focus of \cite{DECAROLIS201787} was on implicit feedback in CRS, where this feedback was obtained from non-verbal communication acts. To assess the effectiveness of using such signals, the accuracy of rating predictions by a content-based recommender was evaluated using MAE and RMSE. In their approach, the ground truth for the evaluation was previously collected in a user study. In some ways, this approach is similar to \cite{Narducci2018aiai} in that the effects of the performance of a side task---here, the interpretation of non-verbal communication acts---on the system's overall recommendation quality are investigated.

Given the possible limitations of pure offline experiments in the context of CRS, user studies are also frequently applied to gauge the effectiveness of a system. In the context of a critiquing-based system \cite{DBLP:conf/aaai/ChenP06,Chen:iui2017}, for example, \emph{decision accuracy} was objectively measured by the fraction of users who changed their mind when they were presented with all available options after they had previously made their selection with the help of the CRS. In \cite{Mahmood:2009:IRS:1557914.1557930}, in contrast, the authors used \emph{task completion rates} and \emph{add-to-cart actions} as proxies, which measure how often users had at least one item in their cart and how many items they added on average, respectively.

\paragraph{Subjective Measures}
Differently from objective measures that, e.g, record the user's decision behavior when interacting with the system or determine prediction accuracy using simulations, subjective measures assess the user's quality perception of the system. Such measurements can be important because even common accuracy measures are often not predictive of the quality of the recommendations as perceived by the users.\footnote{See, e.g., \cite{Rossetti:2016:COO:2959100.2959176,Cremonesi:2012:IPP:2209310.2209314,
Garcin:2014:OOE:2645710.2645745,Maksai:2015:POP:2792838.2800184,DBLP:conf/ercimdl/BeelL15}.} In the reviewed literature on CRS, various quality factors were examined that are also commonly used for non-conversational recommenders, e.g., those discussed in the evaluation frameworks in \cite{umuai2012knijnenburg} and \cite{Pu:2011:UEF:2043932.2043962}.

For the critiquing-based systems discussed in \cite{DBLP:conf/aaai/ChenP06,Chen:iui2017}, the authors therefore not only used decision accuracy (as an objective measure) but also assessed the different factors like \emph{decision confidence}, and \emph{purchase and return intentions}.  \emph{User satisfaction}, either with the system's recommendations or the system as a whole, was additionally investigated in earlier critiquing approaches such as \cite{DBLP:conf/recsys/PuZC09,MapBasedRicci2010} and in other comparative evaluations \cite{Narducci2018aiai,Yang:2018:UUI:3240323.3240389}.
The \emph{perceived recommendation quality} was assessed in the speech-controlled critiquing system in \cite{Grasch:2013:RTC:2507157.2507161}, and in \cite{ResearchNoteContingencyApproach2013} the authors looked at \emph{user acceptance rates}. In \cite{Jin:2019:MEC:3357384.3357923,DBLP:conf/chi/LoeppHZ14} and \cite{Pecune:HAI2019}, finally, the authors considered several dimensions in their questionnaire like the \emph{match of the recommendations} with the preferences (\emph{interest fit}), the \emph{confidence} that the recommendations will be liked, and \emph{trust}.

\subsubsection{Efficiency of Task Support}
Traditionally, in particular critiquing-based CRS approaches are often evaluated in terms of the efficiency of the recommendation process. Specifically, one goal of generating dynamic critiques is to minimize the number of required interactions until the user finds the needed item or accepts the recommendation. Such evaluations are often done \emph{offline} with simulated user profiles. One assumption, also in approaches that are not based on critiquing, is that the simulated users act rationally and consistently, i.e., they will not revise their preferences during the process.

Examples of works that measure interaction cycles in critiquing approaches include \cite{McCarthy2004Onthe,mccarthy2006group,DBLP:conf/recsys/ViappianiB09,Mahmood:2009:IRS:1557914.1557930,DBLP:conf/ah/ZhangP06,reilly2007comparison,MapBasedRicci2010,Grasch:2013:RTC:2507157.2507161}.
The \emph{number of required interaction stages} was also one of usually multiple evaluation criteria for chatbot-like applications, e.g., \cite{Ikemoto2019,Narducci2018aiai,Jin:2019:MEC:3357384.3357923,warnestaal2005user}, and a shopping decision-aid in \cite{ResearchNoteContingencyApproach2013}.
In the context of learning-based systems, the number of dialogue turns in a two-stage interaction model was measured in \cite{Sun:2018:CRS:3209978.3210002}. The usage of such measures is however rather uncommon for natural language, learning-based dialogue systems.

Besides the number of interaction stages, \emph{task completion time} is sometimes used as an alternative or complementary way of objectively measuring efficiency, e.g., in \cite{Mahmood:2009:IRS:1557914.1557930,Jin:2019:MEC:3357384.3357923}.
In \cite{iovine2020conversational}, the authors, among other aspects, compared the efficiency of different interaction modes with a chatbot: NLP-based, button-based, and mixed. They measured the number of questions, the interaction time and the time per question in the dialogue. A main outcome of their work was that pure natural language interfaces were leading to less efficient recommendation sessions, in part due to problems of correctly interpreting the natural language utterances.

In the mentioned papers, shorter interaction or task completion times are generally considered favorable. Note however, that in some cases longer sessions are desirable. In particular, longer interaction times might reflect higher user engagement and, as in \cite{Jin:2019:MEC:3357384.3357923}, correspond to a larger number of listened songs in a music application.
In \cite{DBLP:conf/ecis/ChenHC08}, the authors compared a voice-based and visual output system and measured the number of options that were explored by the users. In this context, note that the exploration of more items can, depending on the application, both be a sign that the user found more interesting options to inspect and a sign that the user did \emph{not} find something immediately and had to explore more options. In \cite{Yang:2018:UUI:3240323.3240389}, the effects of using a voice interface for a podcast recommender were analyzed. Their results showed that users were slower, explored fewer options, and chose fewer long-tail items, which can be detrimental for discovery.

In some works, finally, subjective measures regarding the efficiency of the process are used, typically as a part of usability assessments. In \cite{ResearchNoteContingencyApproach2013,Pecune:HAI2019,DBLP:conf/chi/LoeppHZ14,DBLP:conf/aaai/ChenP06} and \cite{Mahmood:2009:IRS:1557914.1557930} the authors ask the study participants about their perceived \emph{cognitive effort}.

\subsubsection{Quality of the Conversation and Usability Aspects}
In a number of works, the focus of the evaluation is put on certain aspects of the dialogue quality and on usability aspects regarding the system as a whole. The general \emph{ease-of-use} of the system was, for example examined in \cite{Jin:2019:MEC:3357384.3357923,reilly2007comparison,DBLP:conf/recsys/PuZC09,Grasch:2013:RTC:2507157.2507161}; the more specific concept \emph{task ease} was part of the user questionnaire in \cite{warnestaal2005user}.

Regarding quality aspects of the conversation itself, various aspects are investigated in the literature. From the perspective of the conversation initiative, the authors of \cite{DBLP:conf/chi/LoeppHZ14} and \cite{Pecune:HAI2019} measured the perceived level of \emph{user control}. Whether or not the desire for control is dependent on personal characteristics was investigated in \cite{Jin:2019:MEC:3357384.3357923}. In addition to user control, perceived \emph{transparency} was considered as a quality factor in \cite{Pecune:HAI2019}. A common way to establish transparency is through the use of explanations. Questions of how to design explanations for a recommender chatbot were investigated in \cite{Park2019}.
The quality factors used in \cite{warnestaal2005user} were based on an early framework for evaluating spoken dialogue systems in \cite{Litman:EEA:1999}. They, for example, include \emph{adaptation} (i.e., how fast the system adapts to the user's preferences), \emph{expected behavior} (i.e., how intuitive and natural the dialogue interaction is), or the \emph{entertainment} value. Furthermore, in \cite{Pecune:HAI2019} \emph{coordination}, \emph{mutual attentiveness}, \emph{positivity}, and \emph{rapport} were considered as additional desired factors of a conversation.

Looking closer at the content and linguistic level of the dialogues, many recent proposals based on natural language rely on the \emph{BLEU} \cite{BLEU2002} score to assess the system's responses, e.g., \cite{li2017end-to-end,liao2019deep,li2018towards,kang2019recommendation,nie2019multimodal}. With the help of this score, which was developed in the context of machine translation, one can compare the responses generated by the system with ground-truth responses from real human conversations in an automated way. As an alternative, the \emph{NIST} score can be used, e.g., in \cite{nie2019multimodal}. Additional objective linguistic aspects that are measured in the literate include the \emph{lexical diversity} \cite{DBLP:conf/aaai/GhazvininejadBC18}, \emph{perplexity} (corresponding to fluency), and \emph{distinct n-gram} (to assess diversity) \cite{chen-etal-2019-towards}.
In addition to these objective linguistic measures, researchers sometimes consider subjective assessments of the quality of the system responses in their evaluations, e.g., with respect to \emph{fluency}, \emph{appropriateness}, \emph{consistency}, \emph{engagingness}, \emph{relevance}, \emph{informativeness}, and the \emph{overall dialogue quality} and \emph{generation performance} \cite{DBLP:conf/aaai/GhazvininejadBC18,chen-etal-2019-towards,liao2019deep,kang2019recommendation,nie2019multimodal,li2017end-to-end,warnestaal2005user}.

\subsubsection{Effectiveness of Sub Task}
In some works, finally, researchers focus on the evaluation of the performance of certain subtasks. Again, such measurements can both be objective or subjective ones. As objective measurements, the \emph{reward} is often computed in approaches that rely on reinforcement learning \cite{Mahmood:2009:IRS:1557914.1557930}.
In a critiquing system, the number of times a proposed critique was applied was investigated in \cite{reilly2007comparison}. In NLP-based systems, in contrast, researchers often evaluate the performance of the entity and intent recognition modules \cite{liao2019deep,Narducci2018aiai}. In the particular multi-modal CRS in \cite{nie2019multimodal}, Recall was used for assessing the image selection performance. In terms of subjective measures, the \emph{interpretation performance}, i.e., how good the system is in understanding the input, was, for example, considered in \cite{warnestaal2005user}.

\subsection{Discussion}
Our review shows that a wide range of different evaluation methodologies and metrics are used to evaluate CRS. In principle, general user-centric evaluation frameworks for recommender systems as proposed in \cite{umuai2012knijnenburg} and \cite{Pu:2011:UEF:2043932.2043962}  can be applied for CRS as well. So far, however, while user-centric evaluation is common, these frameworks are not widely used and no standards or extensions to them were proposed in the literature. In terms of objective measurements, typical accuracy measures are used by several researchers. Still, the individual CRS proposals in the literature are quite diverse, e.g., in terms of the application domain, interaction strategy, and background knowledge, and a comparison between existing systems remains challenging.

In NLP-based systems, the BLEU score is widely used for automatic evaluation. However, according to  \cite{liu-etal-2016-evaluate}, the BLEU score, at least at the sentence level, can correlate poorly with user perceptions, see also \cite{DBLP:conf/aaai/GhazvininejadBC18}. In general, the evaluation of language models is often considered difficult \cite{MarcusGPT2020} and task-oriented systems like CRS might be even more challenging to assess. These observations therefore suggest that BLEU scores alone cannot inform us well about the quality of the generated system utterances and that in addition subjective evaluations should be applied.

Researchers therefore often resort to offline experiments with simulated users or user studies, where study participants have to accomplish a certain task. In offline studies, often a target (preferred) item is randomly selected, and then a rationally-behaving user is simulated, which interacts with the CRS by answering questions about preferences or by providing feedback on explanations. Such a design however assumes that users \emph{a priori} have fixed preferences towards items or item features. However, in reality, users may also construct or change their preference during the conversation when they learn about the space of options. Therefore, it is not always fully clear to what extent such simulations reflect real-world situations. In user studies, in contrast, often realistic decision situations are explored and participants have to accomplish tasks like selecting a product in a shop or finding musical tracks for a birthday party. While such studies to some extent remain artificial as usually no real purchase is made, such evaluations seem more realistic than the offline experiments described above.  In general, relying solely on offline experimentation seems too limited, except for certain subtasks, given that any CRS is a system that has to support complex user interactions.

Finally, more research seems needed with respect to understand
\begin{enumerate*}[label=\textit{(\roman*)}]
\item how humans make recommendations to each other in a conversation, and \item how users interact with intelligent assistants, e.g., what kind of intelligence they attribute to them and what their expectations are.
\end{enumerate*}
Some aspects related to these questions are discussed, e.g., in \cite{Park2019,kang2017understanding,Yang:2018:UUI:3240323.3240389,christakopoulou2016towards}.
With respect to how humans talk with each other, some analyses were done in \cite{DBLP:conf/ewcbr/Bridge02} and \cite{christakopoulou2016towards}. In \cite{DBLP:conf/ewcbr/Bridge02}, the authors based their research on insights from the field of \emph{Conversational Analysis}, and correspondingly implement typical real-world conversation patterns, albeit in a somewhat restricted form, in their technical proposal.
In general, more work also needs to be done to understand the effects of the quality perception of a system when certain communication patterns like the explanation for a system recommendation are not supported, as it is the case for many investigated systems.

\section{Outlook}
Our study reveals an increased rise in the area of CRS in the past few years, where the most recent approaches rely on machine learning techniques, in particular deep learning, and natural language based interactions.
Despite these advances, a number of research questions remain open, as outlined in the discussion sections throughout the paper. In this final section, we briefly discuss four more  general research directions.

One first question is ``Which interaction modality supports the user best in a given task?''. While voice and written natural language have become more popular recently, more research is required to understand which modality is suited for a given task and situation at hand or if alternative modalities should be offered to the user.
An interesting direction of research also lies in the interpretation of non-verbal communication acts by users.
Furthermore, entirely voice-based CRS have limitations  when it comes to present an entire set of recommendations in one interaction cycle. In such a setting, a \emph{summarization} of a set of recommendations might be needed, as it might in most cases not be meaningful when the CRS reads out several options to the user.

Second, we ask: ``What are challenges and requirements in non-standard application environments?''
Today, most existing research focuses on interactive web or mobile applications, either with forms and buttons or with natural language input in chatbot applications. Some of the discussed works go beyond such scenarios and consider alternative environments where CRS can be used, e.g., within physical stores, in cars, on kiosk solutions, or as a feature of (humanoid) robots. However, little is known so far about the specific requirements, challenges, and opportunities that come with such application scenarios and regarding the critical factors that determine the adoption and value of such systems. Regarding the usage scenarios, most research works discussed in our survey focus on one-to-one communications. However, there are additional scenarios which are not much explored yet, for example, where the CRS supports group decision processes \cite{Alvarez2016,nguyen2017chat}.

A third question is ``What can we learn from theories of conversation?'', see also \cite{thomas2020theories}.
Regarding the \emph{underpinnings} and \emph{adoption factors} of CRS, only very few works are based on concepts and insights from Conversation Analysis, Communication Theory or related fields. In some works, at least certain communication patterns in real-world recommendation dialogues were discussed at a qualitative or anecdotal level. What seems to be mostly missing so far, however, is a clearer understanding of what makes a CRS truly helpful, what users expect from such a system, what makes them fail \cite{BENMIMOUN2012605}, and which intents we should or must support in a system. Explanations are often considered as a main feature for a convincing dialogue, but these aspects are not explored a lot. In addition, more research is required to understand the mechanisms that increase the adoption of CRS, e.g., by increasing the user's trust and developing intimacy \cite{LEE201795}, or by adapting the communication style (e.g., with respect to the initiative and language) to the individual user.

Finally, from a \emph{technical} and \emph{methodological} perspective, we ask: ``How far do we get with pure end-to-end learning approaches, i.e., by creating systems where, besides the item database, only a corpus of past conversations serves as input. Tremendous advances were made in NLP technology in recent years, but it stands to question if today's learning-based CRS are actually useful, see \cite{JannachManzoor2020}. In part, the problem of assessing this aspect is tied to how we evaluate such systems. Computational metrics like BLEU can only answer certain aspects of the question. But also the human evaluations in the reviewed papers are sometimes not too insightful, in particular when a newly proposed system is evaluated relative to a previous system by a few human judges. We therefore should revisit our evaluation practice and also investigate what users actually expect from a CRS, how tolerant they are with respect to misunderstandings or poor recommendations, how we can influence these expectations, and how useful the systems are considered on an absolute scale. Technically, combining learning techniques with other sorts of structured knowledge seems to be key to more usable, reliable and also predictable conversational recommender systems in the future.

\bibliography{acmart}

\begin{thebibliography}{174}
\providecommand{\natexlab}[1]{#1}
\providecommand{\url}[1]{\texttt{#1}}
\expandafter\ifx\csname urlstyle\endcsname\relax
  \providecommand{\doi}[1]{doi: #1}\else
  \providecommand{\doi}{doi: \begingroup \urlstyle{rm}\Url}\fi

\bibitem[{\'A}lvarez~M{\'a}rquez and Ziegler(2016)]{Alvarez2016}
J.~O. {\'A}lvarez~M{\'a}rquez and J.~Ziegler.
\newblock {Hootle+}: {A} group recommender system supporting preference
  negotiation.
\newblock In \emph{Collaboration and Technology}, pages 151--166, 2016.

\bibitem[Andr{\'e} and Pelachaud(2010)]{Andre2010}
E.~Andr{\'e} and C.~Pelachaud.
\newblock Interacting with embodied conversational agents.
\newblock In \emph{Speech Technology: Theory and Applications}, pages 123--149.
  Springer US, 2010.

\bibitem[Angara et~al.(2017)Angara, Jim\'{e}nez, Agarwal, Jain, Jain, Stege,
  Ganti, M\"{u}ller, and Ng]{Angara2017}
P.~Angara, M.~Jim\'{e}nez, K.~Agarwal, H.~Jain, R.~Jain, U.~Stege, S.~Ganti,
  H.~A. M\"{u}ller, and J.~W. Ng.
\newblock {Foodie Fooderson: A Conversational Agent for the Smart Kitchen}.
\newblock In \emph{CASCON '17}, page 247–253, 2017.

\bibitem[{Argal} et~al.(2018){Argal}, {Gupta}, {Modi}, {Pandey}, {Shim}, and
  {Choo}]{Argal2018}
A.~{Argal}, S.~{Gupta}, A.~{Modi}, P.~{Pandey}, S.~{Shim}, and C.~{Choo}.
\newblock Intelligent travel chatbot for predictive recommendation in {E}cho
  platform.
\newblock In \emph{CCWC'18}, pages 176--183, 2018.

\bibitem[{Arteaga} et~al.(2019){Arteaga}, {Arenas}, {Paz}, {Tupia}, and
  {Bruzza}]{Arteag2019}
D.~{Arteaga}, J.~{Arenas}, F.~{Paz}, M.~{Tupia}, and M.~{Bruzza}.
\newblock Design of information system architecture for the recommendation of
  tourist sites in the city of {M}anta, {E}cuador through a chatbot.
\newblock In \emph{CISTI '19}, pages 1--6, 2019.

\bibitem[Ashktorab et~al.(2019)Ashktorab, Jain, Liao, and
  Weisz]{ashktorab2019resilient}
Z.~Ashktorab, M.~Jain, Q.~V. Liao, and J.~D. Weisz.
\newblock Resilient chatbots: {R}epair strategy preferences for conversational
  breakdowns.
\newblock In \emph{CHI'19}, page 254, 2019.

\bibitem[Averjanova et~al.(2008)Averjanova, Ricci, and
  Nguyen]{AverjanovaMobyRec2008}
O.~Averjanova, F.~Ricci, and Q.~N. Nguyen.
\newblock Map-based interaction with a conversational mobile recommender
  system.
\newblock In \emph{UBICOMM '08}, pages 212--218, 2008.

\bibitem[Bader et~al.(2011)Bader, Siegmund, and Woerndl]{Bader2011}
R.~Bader, O.~Siegmund, and W.~Woerndl.
\newblock A study on user acceptance of proactive {In-Vehicle} recommender
  systems.
\newblock In \emph{AutomotiveUI '11}, page 47–54, 2011.

\bibitem[Becker et~al.(2006)Becker, Blaylock, Gerstenberger,
  Kruijff-Korbayov\'{a}, Korthauer, Pinkal, Pitz, Poller, and
  Schehl]{Becker2006PAIS}
T.~Becker, N.~Blaylock, C.~Gerstenberger, I.~Kruijff-Korbayov\'{a},
  A.~Korthauer, M.~Pinkal, M.~Pitz, P.~Poller, and J.~Schehl.
\newblock Natural and intuitive multimodal dialogue for {In-Car} applications:
  {T}he {SAMMIE} {S}ystem.
\newblock In \emph{ECAI '06}, page 612–616, 2006.

\bibitem[Beel and Langer(2015)]{DBLP:conf/ercimdl/BeelL15}
J.~Beel and S.~Langer.
\newblock A comparison of offline evaluations, online evaluations, and user
  studies in the context of research-paper recommender systems.
\newblock In \emph{TPDL '15}, pages 153--168, 2015.

\bibitem[Blanco and Ricci(2013)]{Blanco:Recsys2013}
H.~Blanco and F.~Ricci.
\newblock Acquiring user profiles from implicit feedback in a conversational
  recommender system.
\newblock In \emph{RecSys '13}, page 307–310, 2013.

\bibitem[Bobrow et~al.(1977)Bobrow, Kaplan, Kay, Norman, Thompson, and
  Winograd]{Bobrow1977GUSAF}
D.~G. Bobrow, R.~M. Kaplan, M.~Kay, D.~A. Norman, H.~S. Thompson, and
  T.~Winograd.
\newblock {GUS, A Frame-Driven Dialog System}.
\newblock \emph{Artificial Intelligence}, 8:\penalty0 155--173, 1977.

\bibitem[Bridge(2002)]{DBLP:conf/ewcbr/Bridge02}
D.~G. Bridge.
\newblock Towards conversational recommender systems: {A} dialogue grammar
  approach.
\newblock In \emph{ECCBR '02}, pages 9--22, September 2002.

\bibitem[Burke(1999)]{Burke:1999:WPS:315149.315486}
R.~Burke.
\newblock The {Wasabi} personal shopper: {A} case-based recommender system.
\newblock In \emph{AAAI '99}, pages 844--849, 1999.

\bibitem[Burke et~al.(1997)Burke, Hammond, and
  Young]{DBLP:journals/expert/BurkeHY97}
R.~D. Burke, K.~J. Hammond, and B.~C. Young.
\newblock The {FindMe} approach to assisted browsing.
\newblock \emph{{IEEE} Expert}, 12\penalty0 (4):\penalty0 32--40, 1997.

\bibitem[Cai and Chen(2019)]{cai2019department}
W.~Cai and L.~Chen.
\newblock Towards a taxonomy of user feedback intents for conversational
  recommendations.
\newblock In \emph{RecSys' 19 Late-Breaking Results}, pages 572--573, 2019.

\bibitem[Carenini et~al.(2003)Carenini, Smith, and
  Poole]{DBLP:conf/iui/CareniniSP03}
G.~Carenini, J.~Smith, and D.~Poole.
\newblock Towards more conversational and collaborative recommender systems.
\newblock In \emph{IUI '03}, pages 12--18, 2003.

\bibitem[Carolis et~al.(2017)Carolis, de~Gemmis, Lops, and
  Palestra]{DECAROLIS201787}
B.~D. Carolis, M.~de~Gemmis, P.~Lops, and G.~Palestra.
\newblock Recognizing users feedback from non-verbal communicative acts in
  conversational recommender systems.
\newblock \emph{Pattern Recognition Letters}, 99:\penalty0 87--95, 2017.

\bibitem[Cassell(2001)]{Cassell:2001:ECA:567363.567368}
J.~Cassell.
\newblock Embodied conversational agents: {R}epresentation and intelligence in
  user interfaces.
\newblock \emph{AI Magazine}, 22\penalty0 (4):\penalty0 67--83, 2001.

\bibitem[Chakraborty et~al.(2019)Chakraborty, Anagha, Vats, Baradia, Khan,
  Sarkar, and Roychowdhury]{Chakraborty2019RecommendenceAF}
S.~R. Chakraborty, M.~Anagha, K.~Vats, K.~Baradia, T.~Khan, S.~Sarkar, and
  S.~Roychowdhury.
\newblock Recommendence and fashionsence: {O}nline fashion advisor for offline
  experience.
\newblock In \emph{CoDS-COMAD '19}, 2019.

\bibitem[{Chandrashekara} et~al.(2018){Chandrashekara}, {Talluri},
  {Sivarathri}, {Mitra}, {Calyam}, {Kee}, and {Nair}]{FuzzyBased2018}
A.~A. {Chandrashekara}, R.~K.~M. {Talluri}, S.~S. {Sivarathri}, R.~{Mitra},
  P.~{Calyam}, K.~{Kee}, and S.~{Nair}.
\newblock Fuzzy-based conversational recommender for data-intensive science
  gateway applications.
\newblock In \emph{BigData '18}, pages 4870--4875, 2018.

\bibitem[Chen et~al.(2010)Chen, Jonsson, Villing, and Larsson]{Chen2010}
F.~Chen, I.-M. Jonsson, J.~Villing, and S.~Larsson.
\newblock Application of speech technology in vehicles.
\newblock In \emph{Speech Technology: Theory and Applications}, pages 195--219.
  Springer, 2010.

\bibitem[Chen and Pu(2006)]{DBLP:conf/aaai/ChenP06}
L.~Chen and P.~Pu.
\newblock Evaluating critiquing-based recommender agents.
\newblock In \emph{AAAI '06}, pages 157--162, July 2006.

\bibitem[Chen and Pu(2007{\natexlab{a}})]{Chen:iui2017}
L.~Chen and P.~Pu.
\newblock Hybrid critiquing-based recommender systems.
\newblock In \emph{IUI '07}, page 22–31, 2007{\natexlab{a}}.

\bibitem[Chen and Pu(2007{\natexlab{b}})]{chen2007preference}
L.~Chen and P.~Pu.
\newblock Preference-based organization interfaces: {A}iding user critiques in
  recommender systems.
\newblock In \emph{International Conference on User Modeling}, pages 77--86,
  2007{\natexlab{b}}.

\bibitem[Chen and Pu(2012)]{chenpucritiquing2012}
L.~Chen and P.~Pu.
\newblock Critiquing-based recommenders: survey and emerging trends.
\newblock \emph{User Modeling and User-Adapted Interaction}, 22\penalty0
  (1-2):\penalty0 125--150, 2012.

\bibitem[Chen et~al.(2019)Chen, Lin, Zhang, Ding, Cen, Yang, and
  Tang]{chen-etal-2019-towards}
Q.~Chen, J.~Lin, Y.~Zhang, M.~Ding, Y.~Cen, H.~Yang, and J.~Tang.
\newblock Towards knowledge-based recommender dialog system.
\newblock In \emph{EMNLP-IJCNLP '19}, pages 1803--1813, 2019.

\bibitem[Chen et~al.(2008)Chen, Huang, and Chou]{DBLP:conf/ecis/ChenHC08}
W.~Chen, H.~Huang, and S.~T. Chou.
\newblock Understanding consumer recommendation behavior in a mobile phone
  service context.
\newblock In \emph{ECIS '08}, pages 1022--1033, 2008.

\bibitem[Christakopoulou et~al.(2016)Christakopoulou, Radlinski, and
  Hofmann]{christakopoulou2016towards}
K.~Christakopoulou, F.~Radlinski, and K.~Hofmann.
\newblock Towards conversational recommender systems.
\newblock In \emph{KDD '16}, pages 815--824, 2016.

\bibitem[Christakopoulou et~al.(2018)Christakopoulou, Beutel, Li, Jain, and
  Chi]{Christakopoulou:2018}
K.~Christakopoulou, A.~Beutel, R.~Li, S.~Jain, and E.~H. Chi.
\newblock Q\&{R}: A two-stage approach toward interactive recommendation.
\newblock In \emph{KDD '18}, pages 139--148, 2018.

\bibitem[{Clarizia} et~al.(2018){Clarizia}, {Colace}, {Lombardi}, and
  {Pascale}]{Clarizia2018}
F.~{Clarizia}, F.~{Colace}, M.~{Lombardi}, and F.~{Pascale}.
\newblock A context aware recommender system for digital storytelling.
\newblock In \emph{AINA '18}, pages 542--549, 2018.

\bibitem[Colace et~al.(2017)Colace, De~Santo, Pascale, Lemma, and
  Lombardi]{BotWheels2017}
F.~Colace, M.~De~Santo, F.~Pascale, S.~Lemma, and M.~Lombardi.
\newblock {BotWheels}: {A} petri net based chatbot for recommending tires.
\newblock In \emph{DATA '17}, pages 350--358, 2017.

\bibitem[Contreras et~al.(2015)Contreras, Salam{\'o}, Rodr{\'i}guez, and
  Puig]{Contreras2015A3V}
D.~Contreras, M.~Salam{\'o}, I.~Rodr{\'i}guez, and A.~Puig.
\newblock A {3D} visual interface for critiquing-based recommenders:
  {A}rchitecture and interaction.
\newblock \emph{IJIMAI}, 3:\penalty0 7--15, 2015.

\bibitem[{Contreras} et~al.(2018){Contreras}, {Salamo}, {Rodriguez}, and
  {Puig}]{Contreras2018}
D.~{Contreras}, M.~{Salamo}, I.~{Rodriguez}, and A.~{Puig}.
\newblock Shopping decisions made in a virtual world: {D}efining a state-based
  model of collaborative and conversational user-recommender interactions.
\newblock \emph{IEEE Consumer Electronics Magazine}, 7\penalty0 (4):\penalty0
  260--35, 2018.

\bibitem[Cremonesi et~al.(2012)Cremonesi, Garzotto, and
  Turrin]{Cremonesi:2012:IPP:2209310.2209314}
P.~Cremonesi, F.~Garzotto, and R.~Turrin.
\newblock Investigating the persuasion potential of recommender systems from a
  quality perspective: An empirical study.
\newblock \emph{Transactions on Interactive Intelligent Systems}, 2\penalty0
  (2):\penalty0 1--41, 2012.

\bibitem[Dalton et~al.(2018)Dalton, Ajayi, and
  Main]{Dalton:2018:VGC:3209978.3210168}
J.~Dalton, V.~Ajayi, and R.~Main.
\newblock {Vote Goat}: {C}onversational movie recommendation.
\newblock In \emph{SIGIR '18}, pages 1285--1288, 2018.

\bibitem[De~Carolis et~al.(2015)De~Carolis, de~Gemmis, and
  Lops]{decarolis2015empire}
B.~De~Carolis, M.~de~Gemmis, and P.~Lops.
\newblock A multimodal framework for recognizing emotional feedback in
  conversational recommender systems.
\newblock In \emph{EMPIRE Workshop at ACM RecSys}, page 11–18, 2015.

\bibitem[Dehn and van Mulken(2000)]{DEHN20001}
D.~M. Dehn and S.~van Mulken.
\newblock The impact of animated interface agents: {A} review of empirical
  research.
\newblock \emph{International Journal of Human-Computer Studies}, 52\penalty0
  (1):\penalty0 1 -- 22, 2000.

\bibitem[Dietz et~al.(2019)Dietz, Myftija, and W{\"o}rndl]{Dietz2019}
L.~W. Dietz, S.~Myftija, and W.~W{\"o}rndl.
\newblock Designing a conversational travel recommender system based on
  data-driven destination characterization.
\newblock In \emph{ACM RecSys Workshop on Recommenders in Tourism}, pages
  17--21, 2019.

\bibitem[Dodge et~al.(2016)Dodge, Gane, Zhang, Bordes, Chopra, Miller, Szlam,
  and Weston]{DBLP:journals/corr/DodgeGZBCMSW15}
J.~Dodge, A.~Gane, X.~Zhang, A.~Bordes, S.~Chopra, A.~H. Miller, A.~Szlam, and
  J.~Weston.
\newblock Evaluating prerequisite qualities for learning end-to-end dialog
  systems.
\newblock In \emph{ICLR'16}, 2016.

\bibitem[Felfernig et~al.(2006)Felfernig, Friedrich, Jannach, and
  Zanker]{Felfernig:2006:IED:1278152.1278154}
A.~Felfernig, G.~Friedrich, D.~Jannach, and M.~Zanker.
\newblock An integrated environment for the development of knowledge-based
  recommender applications.
\newblock \emph{International {J}ournal of {E}lectronic {C}ommerce},
  11\penalty0 (2):\penalty0 11--34, 2006.

\bibitem[Felfernig et~al.(2015)Felfernig, Friedrich, Jannach, and
  Zanker]{recsyshandbook2015constraints}
A.~Felfernig, G.~Friedrich, D.~Jannach, and M.~Zanker.
\newblock Constraint-based recommender systems.
\newblock In \emph{Recommender Systems Handbook}, volume~1, pages 161--190.
  Springer, 2015.

\bibitem[Foster and Oberlander(2010)]{Foster2010umuai}
M.~E. Foster and J.~Oberlander.
\newblock User preferences can drive facial expressions: {E}valuating an
  embodied conversational agent in a recommender dialogue system.
\newblock \emph{User {M}odeling and {User-Adapted} {I}nteraction}, 20\penalty0
  (4):\penalty0 341–381, 2010.

\bibitem[Garcin et~al.(2014)Garcin, Faltings, Donatsch, Alazzawi, Bruttin, and
  Huber]{Garcin:2014:OOE:2645710.2645745}
F.~Garcin, B.~Faltings, O.~Donatsch, A.~Alazzawi, C.~Bruttin, and A.~Huber.
\newblock Offline and online evaluation of news recommender systems at
  swissinfo.ch.
\newblock In \emph{RecSys '14}, 2014.

\bibitem[Gedikli et~al.(2014)Gedikli, Jannach, and Ge]{Gedikli2014367}
F.~Gedikli, D.~Jannach, and M.~Ge.
\newblock How should {I} explain? {A} comparison of different explanation types
  for recommender systems.
\newblock \emph{International Journal of Human-Computer Studies}, 72\penalty0
  (4):\penalty0 367 -- 382, 2014.

\bibitem[Ghazvininejad et~al.(2018)Ghazvininejad, Brockett, Chang, Dolan, Gao,
  Yih, and Galley]{DBLP:conf/aaai/GhazvininejadBC18}
M.~Ghazvininejad, C.~Brockett, M.~Chang, B.~Dolan, J.~Gao, W.~Yih, and
  M.~Galley.
\newblock A {Knowledge-Grounded} neural conversation model.
\newblock In \emph{AAAI'18}, pages 5110--5117, 2018.

\bibitem[Grasch et~al.(2013)Grasch, Felfernig, and
  Reinfrank]{Grasch:2013:RTC:2507157.2507161}
P.~Grasch, A.~Felfernig, and F.~Reinfrank.
\newblock {ReCommen}t: {T}owards critiquing-based recommendation with speech
  interaction.
\newblock In \emph{RecSys '13}, pages 157--164, 2013.

\bibitem[Greco et~al.(2017)Greco, Suglia, Basile, and Semeraro]{Greco2017}
C.~Greco, A.~Suglia, P.~Basile, and G.~Semeraro.
\newblock {Converse-Et-Impera}: {E}xploiting deep learning and hierarchical
  reinforcement learning for conversational recommender systems.
\newblock In \emph{AI*IA 2017}, pages 372--386, 2017.

\bibitem[Hammond et~al.(1994)Hammond, Burke, and Schmitt]{hammond1994findme}
K.~J. Hammond, R.~Burke, and K.~Schmitt.
\newblock Case-based approach to knowledge navigation.
\newblock In \emph{AAAI '94}, 1994.

\bibitem[Hariri et~al.(2014)Hariri, Mobasher, and
  Burke]{Hariri:2014:CAI:2645710.2645753}
N.~Hariri, B.~Mobasher, and R.~Burke.
\newblock Context adaptation in interactive recommender systems.
\newblock In \emph{RecSys '14}, pages 41--48, 2014.

\bibitem[Herlocker et~al.(2000)Herlocker, Konstan, and
  Riedl]{Herlocker:2000:ECF:358916.358995}
J.~L. Herlocker, J.~A. Konstan, and J.~Riedl.
\newblock Explaining collaborative filtering recommendations.
\newblock In \emph{CSCW '00}, pages 241--250, 2000.

\bibitem[Hong et~al.(2010)Hong, Huang, Chin, Yen, and
  Lin]{Hong:2010:IAS:2108616.2108681}
Z.-W. Hong, R.-T. Huang, K.-Y. Chin, C.-C. Yen, and J.-M. Lin.
\newblock An interactive agent system for supporting knowledge-based
  recommendation: {A} case study on an {e-Novel} recommender system.
\newblock In \emph{ICUIMC'10}, pages 53:1--53:8, 2010.

\bibitem[Ikemoto et~al.(2019)Ikemoto, Asawavetvutt, Kuwabara, and
  Huang]{Ikemoto2019}
Y.~Ikemoto, V.~Asawavetvutt, K.~Kuwabara, and H.-H. Huang.
\newblock Tuning a conversation strategy for interactive recommendations in a
  chatbot setting.
\newblock \emph{Journal of Information and Telecommunication}, 3\penalty0
  (2):\penalty0 180--195, 2019.

\bibitem[Iovine et~al.(2020)Iovine, Narducci, and
  Semeraro]{iovine2020conversational}
A.~Iovine, F.~Narducci, and G.~Semeraro.
\newblock Conversational recommender systems and natural language: A study
  through the {ConveRSE} framework.
\newblock \emph{Decision Support Systems}, 131:\penalty0 113250--113260, 2020.

\bibitem[Jannach(2004)]{Jannach:2004:ASK:3000001.3000153}
D.~Jannach.
\newblock {ADVISOR SUITE} -- {A} knowledge-based sales advisory system.
\newblock In \emph{ECAI '04}, pages 720--724, 2004.

\bibitem[Jannach(2006)]{Jannach2006}
D.~Jannach.
\newblock Finding preferred query relaxations in content-based recommenders.
\newblock In \emph{IS '06}, pages 355--360, 2006.

\bibitem[Jannach and Jugovac(2019)]{jannachjugovactmis2019}
D.~Jannach and M.~Jugovac.
\newblock Measuring the business value of recommender systems.
\newblock \emph{ACM TMIS}, 10\penalty0 (4):\penalty0 1--23, 2019.

\bibitem[Jannach and Kreutler(2007)]{Jannach:2007:RDK:2011243.2011249}
D.~Jannach and G.~Kreutler.
\newblock Rapid development of knowledge-based conversational recommender
  applications with {Advisor Suite}.
\newblock \emph{Journal of Web Engineering}, 6\penalty0 (2):\penalty0 165--192,
  June 2007.

\bibitem[Jannach and Manzoor(2020)]{JannachManzoor2020}
D.~Jannach and A.~Manzoor.
\newblock End-to-end learning for conversational recommendation: A long way to
  go?
\newblock In \emph{IntRS Workshop at ACM RecSys 2020}, Online, 2020.

\bibitem[Jannach et~al.(2007)Jannach, Zanker, Jessenitschnig, and
  Seidler]{JannachENTER2007}
D.~Jannach, M.~Zanker, M.~Jessenitschnig, and O.~Seidler.
\newblock Developing a conversational travel advisor with {ADVISOR SUITE}.
\newblock In \emph{ENTER '07}, pages 43--52, 2007.

\bibitem[Jannach et~al.(2016)Jannach, Naveed, and
  Jugovac]{JannachNaveedEtAl2016}
D.~Jannach, S.~Naveed, and M.~Jugovac.
\newblock User control in recommender systems: {O}verview and interaction
  challenges.
\newblock In \emph{EC-Web '16}, 2016.

\bibitem[Jin et~al.(2019)Jin, Cai, Chen, Htun, and
  Verbert]{Jin:2019:MEC:3357384.3357923}
Y.~Jin, W.~Cai, L.~Chen, N.~N. Htun, and K.~Verbert.
\newblock {MusicBot}: {E}valuating critiquing-based music recommenders with
  conversational interaction.
\newblock In \emph{CIKM '19}, pages 951--960, 2019.

\bibitem[Joshi et~al.(2017)Joshi, Mi, and
  Faltings]{DBLP:journals/corr/JoshiMF17}
C.~K. Joshi, F.~Mi, and B.~Faltings.
\newblock Personalization in goal-oriented dialog.
\newblock In \emph{NeurIPS '17 Workshop on Conversational AI}, 2017.

\bibitem[Kang et~al.(2019)Kang, Balakrishnan, Shah, Crook, Boureau, and
  Weston]{kang2019recommendation}
D.~Kang, A.~Balakrishnan, P.~Shah, P.~Crook, Y.-L. Boureau, and J.~Weston.
\newblock Recommendation as a communication game: Self-supervised bot-play for
  goal-oriented dialogue.
\newblock In \emph{EMNLP-IJCNLP '19}, pages 1951--1961, Nov. 2019.

\bibitem[Kang et~al.(2017)Kang, Condiff, Chang, Konstan, Terveen, and
  Harper]{kang2017understanding}
J.~Kang, K.~Condiff, S.~Chang, J.~A. Konstan, L.~Terveen, and F.~M. Harper.
\newblock Understanding how people use natural language to ask for
  recommendations.
\newblock In \emph{RecSys '17}, pages 229--237, 2017.

\bibitem[Kapetanios et~al.(2013)Kapetanios, Tatar, and
  Sacarea]{Kapetanios:2013NLP}
E.~Kapetanios, D.~Tatar, and C.~Sacarea.
\newblock \emph{Natural Language Processing: Semantic Aspects}.
\newblock CRC Press, 2013.

\bibitem[Keeney and Raiffa(1993)]{keeney1976decisions}
R.~L. Keeney and H.~Raiffa.
\newblock \emph{{Decisions with Multiple Objectives: Preferences and Value
  Trade-Offs}}.
\newblock Cambridge UP, 1993.

\bibitem[Knijnenburg et~al.(2012)Knijnenburg, Willemsen, Gantner, Soncu, and
  Newell]{umuai2012knijnenburg}
B.~Knijnenburg, M.~Willemsen, Z.~Gantner, H.~Soncu, and C.~Newell.
\newblock Explaining the user experience of recommender systems.
\newblock \emph{User Modeling and User-Adapted Interaction}, 22\penalty0
  (4):\penalty0 441--504, 2012.

\bibitem[Lamche et~al.(2014)Lamche, Adig{\"u}zel, and W{\"o}rndl]{Lamche2014}
B.~Lamche, U.~Adig{\"u}zel, and W.~W{\"o}rndl.
\newblock Interactive explanations in mobile shopping recommender systems.
\newblock In \emph{RecSys '19 IntRS workshop}, pages 14--21, 2014.

\bibitem[Lee et~al.(2016)Lee, Ahn, Lee, Ha, and Lee]{lee2016quote}
H.~Lee, Y.~Ahn, H.~Lee, S.~Ha, and S.-g. Lee.
\newblock Quote recommendation in dialogue using deep neural network.
\newblock In \emph{SIGIR '16}, pages 957--960, 2016.

\bibitem[Lee et~al.(2010)Lee, Kielser, Forlizzi, Srinivasa, and
  Rybski]{LEE:HRI2010}
M.~K. Lee, S.~Kielser, J.~Forlizzi, S.~Srinivasa, and P.~Rybski.
\newblock Gracefully mitigating breakdowns in robotic services.
\newblock In \emph{HRI '10}, page 203–210, 2010.

\bibitem[Lee and Choi(2017)]{LEE201795}
S.~Lee and J.~Choi.
\newblock Enhancing user experience with conversational agent for movie
  recommendation: {E}ffects of self-disclosure and reciprocity.
\newblock \emph{International Journal of Human-Computer Studies}, 103:\penalty0
  95 -- 105, 2017.

\bibitem[Lee et~al.(2018)Lee, Moore, Ren, Arar, and
  Jiang]{DBLP:conf/aaai/LeeMRAJ18}
S.~Lee, R.~J. Moore, G.~Ren, R.~Arar, and S.~Jiang.
\newblock Making personalized recommendation through conversation:
  {A}rchitecture design and recommendation methods.
\newblock In \emph{AAAI '18}, pages 727--730, 2018.

\bibitem[Lee et~al.(2011)Lee, Bae, Kwak, and Kim]{Lee:HRI2011}
Y.~Lee, J.-e. Bae, S.~Kwak, and M.~Kim.
\newblock The effect of politeness strategy on human-robot collaborative
  interaction on malfunction of robot vacuum cleaner.
\newblock In \emph{RSS Workshop on HRI}, 2011.

\bibitem[Li et~al.(2018)Li, Kahou, Schulz, Michalski, Charlin, and
  Pal]{li2018towards}
R.~Li, S.~E. Kahou, H.~Schulz, V.~Michalski, L.~Charlin, and C.~Pal.
\newblock Towards deep conversational recommendations.
\newblock In \emph{NIPS '18}, pages 9725--9735, 2018.

\bibitem[Li et~al.(2017)Li, Chen, Li, Gao, and Celikyilmaz]{li2017end-to-end}
X.~Li, Y.-N. Chen, L.~Li, J.~Gao, and A.~Celikyilmaz.
\newblock End-to-end task-completion neural dialogue systems.
\newblock In \emph{IJCNLP '17}, 2017.

\bibitem[Liao et~al.(2019)Liao, Takanobu, Ma, Yang, Huang, and
  Chua]{liao2019deep}
L.~Liao, R.~Takanobu, Y.~Ma, X.~Yang, M.~Huang, and T.-S. Chua.
\newblock Deep conversational recommender in travel.
\newblock \emph{ArXiv}, abs/1907.00710, 2019.

\bibitem[Litman and Pan(1999)]{Litman:EEA:1999}
D.~J. Litman and S.~Pan.
\newblock Empirically evaluating an adaptable spoken dialogue system.
\newblock In \emph{UM '99}, pages 55--64, 1999.

\bibitem[Liu et~al.(2016)Liu, Lowe, Serban, Noseworthy, Charlin, and
  Pineau]{liu-etal-2016-evaluate}
C.-W. Liu, R.~Lowe, I.~Serban, M.~Noseworthy, L.~Charlin, and J.~Pineau.
\newblock How {NOT} to evaluate your dialogue system: An empirical study of
  unsupervised evaluation metrics for dialogue response generation.
\newblock In \emph{EMNLP '16}, pages 2122--2132, 2016.

\bibitem[Liu et~al.(2010)Liu, Seneff, and Zue]{liu2010dialogue}
J.~Liu, S.~Seneff, and V.~Zue.
\newblock Dialogue-oriented review summary generation for spoken dialogue
  recommendation systems.
\newblock In \emph{ACL '10}, pages 64--72, 2010.

\bibitem[Loepp et~al.(2014)Loepp, Hussein, and
  Ziegler]{DBLP:conf/chi/LoeppHZ14}
B.~Loepp, T.~Hussein, and J.~Ziegler.
\newblock Choice-based preference elicitation for collaborative filtering
  recommender systems.
\newblock In \emph{CHI '14}, pages 3085--3094, 2014.

\bibitem[Loh et~al.(2010)Loh, Lichtnow, Kampff, and de~Oliveira]{Stanley2010}
S.~Loh, D.~Lichtnow, A.~J.~C. Kampff, and J.~P.~M. de~Oliveira.
\newblock Recommendation of complementary material during chat discussions.
\newblock \emph{Knowledge Management \& E-Learning}, 2\penalty0 (4), 2010.

\bibitem[Luettin et~al.(2019)Luettin, Rothermel, and Andrew]{Luettin2019}
J.~Luettin, S.~Rothermel, and M.~Andrew.
\newblock Future of {In-Vehicle} recommendation systems @ {B}osch.
\newblock In \emph{RecSys ’19}, page 524, 2019.

\bibitem[Luo et~al.(2020)Luo, Sanner, Wu, Li, and
  Yang]{LuoLatentLinearCritquing20}
K.~Luo, S.~Sanner, G.~Wu, H.~Li, and H.~Yang.
\newblock Latent linear critiquing for conversational recommender systems.
\newblock In \emph{WWW '20}, page 2535–2541, 2020.

\bibitem[Mahmood and Ricci(2007)]{mahmood2007learning}
T.~Mahmood and F.~Ricci.
\newblock Learning and adaptivity in interactive recommender systems.
\newblock In \emph{ICEC '07}, pages 75--84, 2007.

\bibitem[Mahmood and Ricci(2009)]{Mahmood:2009:IRS:1557914.1557930}
T.~Mahmood and F.~Ricci.
\newblock Improving recommender systems with adaptive conversational
  strategies.
\newblock In \emph{HT '09}, pages 73--82, 2009.

\bibitem[Maksai et~al.(2015)Maksai, Garcin, and
  Faltings]{Maksai:2015:POP:2792838.2800184}
A.~Maksai, F.~Garcin, and B.~Faltings.
\newblock Predicting online performance of news recommender systems through
  richer evaluation metrics.
\newblock In \emph{RecSys '15}, pages 179--186, 2015.

\bibitem[Marcus(2020)]{MarcusGPT2020}
G.~Marcus.
\newblock {GPT-2 and the Nature of Intelligence}.
\newblock \url{https://thegradient.pub/gpt2-and-the-nature-of-intelligence/},
  Jan. 2020.

\bibitem[McCarthy et~al.(2004{\natexlab{a}})McCarthy, Reilly, McGinty, and
  Smyth]{McCarthy2004Onthe}
K.~McCarthy, J.~Reilly, L.~McGinty, and B.~Smyth.
\newblock On the dynamic generation of compound critiques in conversational
  recommender systems.
\newblock In \emph{AH '04}, pages 176--184, 2004{\natexlab{a}}.

\bibitem[McCarthy et~al.(2004{\natexlab{b}})McCarthy, Reilly, McGinty, and
  Smyth]{mccarthy2004thinking}
K.~McCarthy, J.~Reilly, L.~McGinty, and B.~Smyth.
\newblock Thinking positively-explanatory feedback for conversational
  recommender systems.
\newblock In \emph{ECCBR '04}, pages 115--124, 2004{\natexlab{b}}.

\bibitem[McCarthy et~al.(2006)McCarthy, Salam{\'o}, Coyle, McGinty, Smyth, and
  Nixon]{mccarthy2006group}
K.~McCarthy, M.~Salam{\'o}, L.~Coyle, L.~McGinty, B.~Smyth, and P.~Nixon.
\newblock Group recommender systems: {A} critiquing based approach.
\newblock In \emph{IUI '06}, pages 267--269, 2006.

\bibitem[McCarthy et~al.(2010)McCarthy, Salem, and
  Smyth]{mccarthy2010experience}
K.~McCarthy, Y.~Salem, and B.~Smyth.
\newblock Experience-based critiquing: Reusing critiquing experiences to
  improve conversational recommendation.
\newblock In \emph{ICCBR '10}, pages 480--494, 2010.

\bibitem[McGinty and Smyth(2003)]{DBLP:conf/iccbr/McGintyS03}
L.~McGinty and B.~Smyth.
\newblock On the role of diversity in conversational recommender systems.
\newblock In \emph{ICCBR '03}, pages 276--290, 2003.
\newblock \doi{10.1007/3-540-45006-8_23}.

\bibitem[McSherry(2004)]{DBLP:conf/ewcbr/McSherry04a}
D.~McSherry.
\newblock Incremental relaxation of unsuccessful queries.
\newblock In \emph{ECCBR '04}, pages 331--345, 2004.

\bibitem[Mimoun et~al.(2012)Mimoun, Poncin, and Garnier]{BENMIMOUN2012605}
M.~S.~B. Mimoun, I.~Poncin, and M.~Garnier.
\newblock Case study--embodied virtual agents: An analysis on reasons for
  failure.
\newblock \emph{Journal of Retailing and Consumer Services}, 19\penalty0
  (6):\penalty0 605 -- 612, 2012.

\bibitem[Mirzadeh et~al.(2005)Mirzadeh, Ricci, and
  Bansal]{DBLP:conf/eee/MirzadehRB05}
N.~Mirzadeh, F.~Ricci, and M.~Bansal.
\newblock Feature selection methods for conversational recommender systems.
\newblock In \emph{EEE '05}, pages 772--777, April 2005.

\bibitem[Moon et~al.(2019)Moon, Shah, Kumar, and Subba]{moon2019opendialkg}
S.~Moon, P.~Shah, A.~Kumar, and R.~Subba.
\newblock {OpenDialKG}: Explainable conversational reasoning with
  attention-based walks over knowledge graphs.
\newblock In \emph{ACL '19}, pages 845--854, 2019.

\bibitem[Myers et~al.(2018)Myers, Furqan, Nebolsky, Caro, and
  Zhu]{Myers:CHI2018}
C.~Myers, A.~Furqan, J.~Nebolsky, K.~Caro, and J.~Zhu.
\newblock Patterns for how users overcome obstacles in voice user interfaces.
\newblock In \emph{CHI '18}, 2018.

\bibitem[Nadeau and Sekine(2007)]{nadeau2007survey}
D.~Nadeau and S.~Sekine.
\newblock A survey of named entity recognition and classification.
\newblock \emph{Lingvisticae Investigationes}, 30\penalty0 (1):\penalty0 3--26,
  2007.

\bibitem[Narducci et~al.(2018{\natexlab{a}})Narducci, Basile, Iovine,
  de~Gemmis, Lops, and Semeraro]{DBLP:conf/recsys/NarducciBIGLS18}
F.~Narducci, P.~Basile, A.~Iovine, M.~de~Gemmis, P.~Lops, and G.~Semeraro.
\newblock A domain-independent framework for building conversational
  recommender systems.
\newblock In \emph{RecSys '18 KaRS Workshop}, pages 29--34, 2018{\natexlab{a}}.

\bibitem[Narducci et~al.(2018{\natexlab{b}})Narducci, de~Gemmis, Lops, and
  Semeraro]{Narducci2018aiai}
F.~Narducci, M.~de~Gemmis, P.~Lops, and G.~Semeraro.
\newblock Improving the user experience with a conversational recommender
  system.
\newblock In \emph{AI*IA '18}, pages 528--538, 2018{\natexlab{b}}.

\bibitem[Narducci et~al.(2019)Narducci, Basile, de~Gemmis, Lops, and
  Semeraro]{narducci2019investigation}
F.~Narducci, P.~Basile, M.~de~Gemmis, P.~Lops, and G.~Semeraro.
\newblock An investigation on the user interaction modes of conversational
  recommender systems for the music domain.
\newblock \emph{UMUAI '19}, pages 1--34, 2019.

\bibitem[Nguyen and Ricci(2017)]{nguyen2017chat}
T.~N. Nguyen and F.~Ricci.
\newblock A chat-based group recommender system for tourism.
\newblock In \emph{ENTER '17}, pages 17--30, 2017.

\bibitem[Nica et~al.(2018)Nica, Tazl, and Wotawa]{nica2018chatbot}
I.~Nica, O.~A. Tazl, and F.~Wotawa.
\newblock Chatbot-based tourist recommendations using model-based reasoning.
\newblock In \emph{Configuration Workshop '18}, pages 25--30, 2018.

\bibitem[Nie et~al.(2019)Nie, Wang, Hong, Wang, and Tian]{nie2019multimodal}
L.~Nie, W.~Wang, R.~Hong, M.~Wang, and Q.~Tian.
\newblock Multimodal dialog system: Generating responses via adaptive decoders.
\newblock In \emph{MM '19}, pages 1098--1106, 2019.

\bibitem[Nunes and Jannach(2017)]{Explanations:UMUAI2017}
I.~Nunes and D.~Jannach.
\newblock A systematic review and taxonomy of explanations in decision support
  and recommender systems.
\newblock \emph{User Modeling and User-Adapted Interaction}, 27\penalty0
  (3–5):\penalty0 393–444, Dec. 2017.

\bibitem[Papineni et~al.(2002)Papineni, Roukos, Ward, and Zhu]{BLEU2002}
K.~Papineni, S.~Roukos, T.~Ward, and W.-J. Zhu.
\newblock {BLEU}: {A} method for automatic evaluation of machine translation.
\newblock In \emph{ACL '02}, page 311–318, 2002.

\bibitem[Park and Youn-kyung(2019)]{Park2019}
S.~Park and L.~Youn-kyung.
\newblock Design considerations for explanations made by a recommender chatbot.
\newblock In \emph{IASDR '19}, 2019.

\bibitem[Pecune et~al.(2019)Pecune, Murali, Tsai, Matsuyama, and
  Cassell]{Pecune:HAI2019}
F.~Pecune, S.~Murali, V.~Tsai, Y.~Matsuyama, and J.~Cassell.
\newblock A model of social explanations for a conversational movie
  recommendation system.
\newblock In \emph{HAI '19}, page 135–143, 2019.

\bibitem[Pu and Chen(2006)]{ExplainA}
P.~Pu and L.~Chen.
\newblock Trust building with explanation interfaces.
\newblock In \emph{IUI '06}, pages 93--100, 2006.

\bibitem[Pu et~al.(2006)Pu, Viappiani, and Faltings]{Pu:CHI2006}
P.~Pu, P.~Viappiani, and B.~Faltings.
\newblock Increasing user decision accuracy using suggestions.
\newblock In \emph{CHI '06}, page 121–130, 2006.

\bibitem[Pu et~al.(2009)Pu, Zhou, and Castagnos]{DBLP:conf/recsys/PuZC09}
P.~Pu, M.~Zhou, and S.~Castagnos.
\newblock Critiquing recommenders for public taste products.
\newblock In \emph{RecSys '09}, pages 249--252, 2009.

\bibitem[Pu et~al.(2011)Pu, Chen, and Hu]{Pu:2011:UEF:2043932.2043962}
P.~Pu, L.~Chen, and R.~Hu.
\newblock A user-centric evaluation framework for recommender systems.
\newblock In \emph{RecSys '11}, pages 157--164, 2011.

\bibitem[Qiu et~al.(2017)Qiu, Li, Wang, Gao, Chen, Zhao, Chen, Huang, and
  Chu]{qiu2017alime}
M.~Qiu, F.-L. Li, S.~Wang, X.~Gao, Y.~Chen, W.~Zhao, H.~Chen, J.~Huang, and
  W.~Chu.
\newblock Alime chat: {A} sequence to sequence and rerank based chatbot engine.
\newblock In \emph{ACL'17}, pages 498--503, 2017.

\bibitem[Radlinski and Craswell(2017)]{Radlinski:2017:TFC:3020165.3020183}
F.~Radlinski and N.~Craswell.
\newblock A theoretical framework for conversational search.
\newblock In \emph{CHIIR '17}, pages 117--126, 2017.

\bibitem[Radlinski et~al.(2019)Radlinski, Balog, Byrne, and
  Krishnamoorthi]{radlinski2019coached}
F.~Radlinski, K.~Balog, B.~Byrne, and K.~Krishnamoorthi.
\newblock Coached conversational preference elicitation: A case study in
  understanding movie preferences.
\newblock \emph{SIGDIAL '19}, 2019.

\bibitem[Rafailidis and
  Manolopoulos(2019{\natexlab{a}})]{Rafailidis:2019:VAP:3326467.3326468}
D.~Rafailidis and Y.~Manolopoulos.
\newblock Can virtual assistants produce recommendations?
\newblock In \emph{WIMS '19}, 2019{\natexlab{a}}.

\bibitem[Rafailidis and
  Manolopoulos(2019{\natexlab{b}})]{rafailidis2018technological}
D.~Rafailidis and Y.~Manolopoulos.
\newblock The technological gap between virtual assistants and recommendation
  systems.
\newblock \emph{ArXiv}, abs/1901.00431, 2019{\natexlab{b}}.

\bibitem[Rana and Bridge(2020)]{ArbpitNavigationByPreference2020}
A.~Rana and D.~Bridge.
\newblock Navigation-by-preference: A new conversational recommender with
  preference-based feedback.
\newblock In \emph{IUI '20}, page 155–165, 2020.

\bibitem[Rashid et~al.(2002)Rashid, Albert, Cosley, Lam, McNee, Konstan, and
  Riedl]{AM:IUI2002}
A.~M. Rashid, I.~Albert, D.~Cosley, S.~K. Lam, S.~M. McNee, J.~A. Konstan, and
  J.~Riedl.
\newblock Getting to know you: {L}earning new user preferences in recommender
  systems.
\newblock In \emph{IUI ’02}, page 127–134, 2002.

\bibitem[Reilly et~al.(2004)Reilly, McCarthy, McGinty, and
  Smyth]{Reilly:ACBR2004}
J.~Reilly, K.~McCarthy, L.~McGinty, and B.~Smyth.
\newblock Dynamic critiquing.
\newblock In \emph{ECCBR 04}, pages 763--777, 2004.

\bibitem[Reilly et~al.(2007)Reilly, Zhang, McGinty, Pu, and
  Smyth]{reilly2007comparison}
J.~Reilly, J.~Zhang, L.~McGinty, P.~Pu, and B.~Smyth.
\newblock A comparison of two compound critiquing systems.
\newblock In \emph{IUI '07}, pages 317--320, 2007.

\bibitem[Ricci and Nguyen(2007)]{DBLP:journals/expert/RicciN07}
F.~Ricci and Q.~N. Nguyen.
\newblock Acquiring and revising preferences in a critique-based mobile
  recommender system.
\newblock \emph{Intelligent Systems}, 22\penalty0 (3):\penalty0 22--29, 2007.

\bibitem[Ricci et~al.(2003)Ricci, Venturini, Cavada, Mirzadeh, Blaas, and
  Nones]{RicciQueryManagement2003}
F.~Ricci, A.~Venturini, D.~Cavada, N.~Mirzadeh, D.~Blaas, and M.~Nones.
\newblock Product recommendation with interactive query management and twofold
  similarity.
\newblock In \emph{ICCBR '03}, pages 479--493, 2003.

\bibitem[Ricci et~al.(2010)Ricci, Nguyen, and Averjanova]{MapBasedRicci2010}
F.~Ricci, Q.~N. Nguyen, and O.~Averjanova.
\newblock Exploiting a map-based interface in conversational recommender
  systems for mobile travelers.
\newblock In \emph{Tourism Informatics}, pages 73--79. IGI, 2010.

\bibitem[Ricci et~al.(2015)Ricci, Rokach, Shapira, and
  Kantor]{Ricci:2010:RSH:1941884}
F.~Ricci, L.~Rokach, B.~Shapira, and P.~B. Kantor.
\newblock \emph{Recommender Systems Handbook}.
\newblock Springer-Verlag, 2nd edition, 2015.

\bibitem[Rich(1979)]{RICH1979329}
E.~Rich.
\newblock User modeling via stereotypes.
\newblock \emph{Cognitive Science}, 3\penalty0 (4), 1979.

\bibitem[Rossetti et~al.(2016)Rossetti, Stella, and
  Zanker]{Rossetti:2016:COO:2959100.2959176}
M.~Rossetti, F.~Stella, and M.~Zanker.
\newblock Contrasting offline and online results when evaluating recommendation
  algorithms.
\newblock In \emph{RecSys '16}, pages 31--34, 2016.

\bibitem[Saha et~al.(2018)Saha, Khapra, and Sankaranarayanan]{Saha:1704.00200}
A.~Saha, M.~M. Khapra, and K.~Sankaranarayanan.
\newblock Towards building large scale multimodal domain-aware conversation
  systems.
\newblock In \emph{AAAI '18}, 2018.

\bibitem[Salem et~al.(2014)Salem, Hong, and Liu]{Yasser:WWW2016:HistoryGuided}
Y.~Salem, J.~Hong, and W.~Liu.
\newblock {History-Guided} conversational recommendation.
\newblock In \emph{WWW ’14}, page 999–1004, 2014.

\bibitem[Shani and Gunawardana(2011)]{Shani2011}
G.~Shani and A.~Gunawardana.
\newblock Evaluating recommendation systems.
\newblock In \emph{Recommender Systems Handbook}, pages 257--297. Springer US,
  2011.

\bibitem[Shimazu(2002)]{Shimazu2002}
H.~Shimazu.
\newblock {ExpertClerk}: A conversational case-based reasoning tool for
  developing salesclerk agents in {E-Commerce} webshops.
\newblock \emph{Artificial Intelligence Review}, 18\penalty0 (3-4):\penalty0
  223--244, 2002.

\bibitem[{Siangchin} and {Samanchuen}(2019)]{SinghcinICD102019}
N.~{Siangchin} and T.~{Samanchuen}.
\newblock Chatbot implementation for {ICD-10} recommendation system.
\newblock In \emph{ICESI '19}, pages 1--6, 2019.

\bibitem[Smyth et~al.(2004)Smyth, McGinty, Reilly, and
  McCarthy]{smyth2004compound}
B.~Smyth, L.~McGinty, J.~Reilly, and K.~McCarthy.
\newblock Compound critiques for conversational recommender systems.
\newblock In \emph{WI '04}, pages 145--151, 2004.

\bibitem[Sordoni et~al.(2015)Sordoni, Bengio, Vahabi, Lioma, Grue~Simonsen, and
  Nie]{Sordoni:CIKM2015}
A.~Sordoni, Y.~Bengio, H.~Vahabi, C.~Lioma, J.~Grue~Simonsen, and J.-Y. Nie.
\newblock A hierarchical recurrent encoder-decoder for generative context-aware
  query suggestion.
\newblock In \emph{CIKM '15}, page 553–562, 2015.

\bibitem[Srinivasan and Takayama(2016)]{Srinivasan:CHI2016}
V.~Srinivasan and L.~Takayama.
\newblock Help me please: {R}obot politeness strategies for soliciting help
  from humans.
\newblock In \emph{CHI '16}, page 4945–4955, 2016.

\bibitem[Stolcke et~al.(2000)Stolcke, Coccaro, Bates, Taylor, Van Ess-Dykema,
  Ries, Shriberg, Jurafsky, Martin, and Meteer]{Stolcke:CL2000}
A.~Stolcke, N.~Coccaro, R.~Bates, P.~Taylor, C.~Van Ess-Dykema, K.~Ries,
  E.~Shriberg, D.~Jurafsky, R.~Martin, and M.~Meteer.
\newblock Dialogue act modeling for automatic tagging and recognition of
  conversational speech.
\newblock \emph{Computational Linguistics}, 26\penalty0 (3):\penalty0 339--373,
  2000.

\bibitem[Sun et~al.(2013)Sun, Li, Lee, Zhou, Lebanon, and
  Zha]{Sun:2013:LMD:2433396.2433451}
M.~Sun, F.~Li, J.~Lee, K.~Zhou, G.~Lebanon, and H.~Zha.
\newblock Learning multiple-question decision trees for cold-start
  recommendation.
\newblock In \emph{WSDM '13}, pages 445--454, 2013.

\bibitem[Sun and Zhang(2018)]{Sun:2018:CRS:3209978.3210002}
Y.~Sun and Y.~Zhang.
\newblock Conversational recommender system.
\newblock In \emph{SIGIR '18}, pages 235--244, 2018.

\bibitem[Telang et~al.(2018)Telang, Kalia, Vukovic, Pandita, and
  Singh]{Telang2018}
P.~R. Telang, A.~K. Kalia, M.~Vukovic, R.~Pandita, and M.~P. Singh.
\newblock A conceptual framework for engineering chatbots.
\newblock \emph{IEEE Internet Computing}, 22\penalty0 (06):\penalty0 54--59,
  2018.

\bibitem[Thomas et~al.(2020)Thomas, Czerwinski, McDuff, and
  Craswell]{thomas2020theories}
P.~Thomas, M.~Czerwinski, D.~McDuff, and N.~Craswell.
\newblock Theories of conversation for conversational ir.
\newblock In \emph{International Workshop on Conversational Approaches to
  Information Retrieval}, 2020.

\bibitem[Thompson et~al.(2004)Thompson, G\"{o}ker, and
  Langley]{Thompson:2004:PSC:1622467.1622479}
C.~A. Thompson, M.~H. G\"{o}ker, and P.~Langley.
\newblock A personalized system for conversational recommendations.
\newblock \emph{Journal of Artificial Intelligence Research}, 21\penalty0
  (1):\penalty0 393--428, 2004.

\bibitem[Tintarev and Masthoff(2011)]{Tintarev2011Expl}
N.~Tintarev and J.~Masthoff.
\newblock Designing and evaluating explanations for recommender systems.
\newblock In \emph{Recommender Systems Handbook}, volume~1, pages 479--510.
  Springer, 2011.

\bibitem[Tou et~al.(1982)Tou, Williams, Fikes, Jr., and
  Malone]{DBLP:conf/aaai/TouWFHM82}
F.~N. Tou, M.~D. Williams, R.~Fikes, D.~A.~H. Jr., and T.~W. Malone.
\newblock {RABBIT}: An intelligent database assistant.
\newblock In \emph{AAAI '82}, pages 314--318, 1982.

\bibitem[Trabelsi et~al.(2010)Trabelsi, Wilson, Bridge, and
  Ricci]{DBLP:conf/ictai/TrabelsiWBR10}
W.~Trabelsi, N.~Wilson, D.~G. Bridge, and F.~Ricci.
\newblock Comparing approaches to preference dominance for conversational
  recommenders.
\newblock In \emph{ICTAI '10}, pages 113--120, October 2010.

\bibitem[Tsumita and Takagi(2019)]{tsumita2019dialogue}
D.~Tsumita and T.~Takagi.
\newblock Dialogue based recommender system that flexibly mixes utterances and
  recommendations.
\newblock In \emph{WI '19}, pages 51--58, 2019.

\bibitem[Viappiani and Boutilier(2009)]{DBLP:conf/recsys/ViappianiB09}
P.~Viappiani and C.~Boutilier.
\newblock Regret-based optimal recommendation sets in conversational
  recommender systems.
\newblock In \emph{RecSys '11}, pages 101--108, 2009.

\bibitem[Viappiani et~al.(2007)Viappiani, Pu, and
  Faltings]{DBLP:conf/recsys/ViappianiPF07}
P.~Viappiani, P.~Pu, and B.~Faltings.
\newblock Conversational recommenders with adaptive suggestions.
\newblock In \emph{RecSys '07}, pages 89--96, 2007.

\bibitem[Vig et~al.(2011)Vig, Sen, and Riedl]{vigTagGenome2011}
J.~Vig, S.~Sen, and J.~Riedl.
\newblock Navigating the tag genome.
\newblock In \emph{IUI '11}, page 93–102, 2011.

\bibitem[Walker et~al.(2004)Walker, Whittaker, Stent, Maloor, Moore, Johnston,
  and Vasireddy]{WALKER2004811}
M.~Walker, S.~Whittaker, A.~Stent, P.~Maloor, J.~Moore, M.~Johnston, and
  G.~Vasireddy.
\newblock Generation and evaluation of user tailored responses in multimodal
  dialogue.
\newblock \emph{Cognitive Science}, 28\penalty0 (5):\penalty0 811 -- 840, 2004.

\bibitem[Wallace(2009)]{WallaceALICE}
R.~S. Wallace.
\newblock {The Anatomy of A.L.I.C.E.}
\newblock In \emph{Parsing the Turing Test}, pages 181--210. Springer, 2009.

\bibitem[Wang and Benbasat(2013)]{ResearchNoteContingencyApproach2013}
W.~Wang and I.~Benbasat.
\newblock Research {Note—A} contingency approach to investigating the effects
  of user-system interaction modes of online decision aids.
\newblock \emph{Information Systems Research}, 24\penalty0 (3):\penalty0
  861--876, 2013.

\bibitem[W{\"a}rnest{\aa}l(2005{\natexlab{a}})]{warnestaal2005modeling}
P.~W{\"a}rnest{\aa}l.
\newblock Modeling a dialogue strategy for personalized movie recommendations.
\newblock In \emph{IUI '05 Beyond Personalization Workshop}, pages 77--82,
  2005{\natexlab{a}}.

\bibitem[W{\"a}rnest{\aa}l(2005{\natexlab{b}})]{warnestaal2005user}
P.~W{\"a}rnest{\aa}l.
\newblock User evaluation of a conversational recommender system.
\newblock In \emph{IJCAI '05 Workshop on Knowledge and Reasoning in Practical
  Dialogue Systems}, 2005{\natexlab{b}}.

\bibitem[W{\"a}rnest{\aa}l et~al.(2007{\natexlab{a}})W{\"a}rnest{\aa}l,
  Degerstedt, and J{\"o}nsson]{warnestaal2007interview}
P.~W{\"a}rnest{\aa}l, L.~Degerstedt, and A.~J{\"o}nsson.
\newblock Interview and delivery: {D}ialogue strategies for conversational
  recommender systems.
\newblock In \emph{NODALIDA '07}, pages 199--205, 2007{\natexlab{a}}.

\bibitem[W{\"a}rnest{\aa}l et~al.(2007{\natexlab{b}})W{\"a}rnest{\aa}l,
  Degerstedt, and J{\"o}nsson]{warnestaal2007pcql}
P.~W{\"a}rnest{\aa}l, L.~Degerstedt, and A.~J{\"o}nsson.
\newblock {PCQL}: A formalism for human-like preference dialogues*.
\newblock In \emph{IJCAI '07 Workshop on Knowledge and Reasoning in Practical
  Dialogue Systems}, 2007{\natexlab{b}}.

\bibitem[Wei et~al.(2015)Wei, Liu, Zheng, Zhang, Wang, and Wu]{WEI201580}
B.~Wei, J.~Liu, Q.~Zheng, W.~Zhang, C.~Wang, and B.~Wu.
\newblock {DF-Miner}: {D}omain-specific facet mining by leveraging the
  hyperlink structure of {W}ikipedia.
\newblock \emph{Knowledge-Based Systems}, 77:\penalty0 80 -- 91, 2015.

\bibitem[Weizenbaum(1966)]{Weizenbaum:1966:ECP:365153.365168}
J.~Weizenbaum.
\newblock {ELIZA} -- {C}omputer program for the study of natural language
  communication between man and machine.
\newblock \emph{Communications. ACM}, 9\penalty0 (1):\penalty0 36--45, Jan.
  1966.

\bibitem[Wen et~al.(2017)Wen, Vandyke, Mrk{\v{s}}i{\'c}, Ga{\v{s}}i{\'c},
  Rojas-Barahona, Su, Ultes, and Young]{wen-etal-2017-network}
T.-H. Wen, D.~Vandyke, N.~Mrk{\v{s}}i{\'c}, M.~Ga{\v{s}}i{\'c}, L.~M.
  Rojas-Barahona, P.-H. Su, S.~Ultes, and S.~Young.
\newblock A network-based end-to-end trainable task-oriented dialogue system.
\newblock In \emph{ACL '17}, pages 438--449, 2017.

\bibitem[Widyantoro and Baizal(2014)]{widyantoro2014framework}
D.~H. Widyantoro and Z.~Baizal.
\newblock A framework of conversational recommender system based on user
  functional requirements.
\newblock In \emph{ICoICT '14}, pages 160--165, 2014.

\bibitem[Wissbroecker and Harper(2018)]{wissbroecker2018early}
J.~Wissbroecker and F.~M. Harper.
\newblock Early lessons from a voice-only interface for finding movies.
\newblock In \emph{RecSys '19 Late-Breaking Results}, 2018.

\bibitem[Wu et~al.(2019)Wu, Madotto, Hosseini-Asl, Xiong, Socher, and
  Fung]{wu2019transferable}
C.-S. Wu, A.~Madotto, E.~Hosseini-Asl, C.~Xiong, R.~Socher, and P.~Fung.
\newblock Transferable multi-domain state generator for task-oriented dialogue
  systems.
\newblock In \emph{ACL}, 2019.

\bibitem[Xu et~al.(2017)Xu, Benbasat, and
  Cenfetelli]{DBLP:journals/jmis/XuBC17}
D.~J. Xu, I.~Benbasat, and R.~T. Cenfetelli.
\newblock A {Two-Stage} model of generating product advice: {P}roposing and
  testing the complementarity principle.
\newblock \emph{Journal of Management Information Systems}, 34\penalty0
  (3):\penalty0 826--862, 2017.

\bibitem[Yan et~al.(2017)Yan, Duan, Chen, Zhou, Zhou, and
  Li]{DBLP:conf/aaai/YanDCZZL17}
Z.~Yan, N.~Duan, P.~Chen, M.~Zhou, J.~Zhou, and Z.~Li.
\newblock Building task-oriented dialogue systems for online shopping.
\newblock In \emph{AAAI '17}, pages 4618--4626, 2017.

\bibitem[Yang et~al.(2018)Yang, Sobolev, Tsangouri, and
  Estrin]{Yang:2018:UUI:3240323.3240389}
L.~Yang, M.~Sobolev, C.~Tsangouri, and D.~Estrin.
\newblock Understanding user interactions with podcast recommendations
  delivered via voice.
\newblock In \emph{RecSys '18}, pages 190--194, 2018.

\bibitem[Yin et~al.(2017)Yin, Chang, and Zhang]{Yin:2017:DID:3097983.3098148}
Z.~Yin, K.-h. Chang, and R.~Zhang.
\newblock {DeepProbe}: {I}nformation directed sequence understanding and
  chatbot design via recurrent neural networks.
\newblock In \emph{KDD '17}, pages 2131--2139, 2017.

\bibitem[Yu et~al.(2019{\natexlab{a}})Yu, Shen, and
  Jin]{Yu:2019:VDA:3292500.3330991}
T.~Yu, Y.~Shen, and H.~Jin.
\newblock A visual dialog augmented interactive recommender system.
\newblock In \emph{KDD '19}, pages 157--165, 2019{\natexlab{a}}.

\bibitem[Yu et~al.(2019{\natexlab{b}})Yu, Shen, Zhang, Zeng, and
  Jin]{Tong:MM2019:VisionLanguage}
T.~Yu, Y.~Shen, R.~Zhang, X.~Zeng, and H.~Jin.
\newblock Vision-language recommendation via attribute augmented multimodal
  reinforcement learning.
\newblock In \emph{MM ’19}, page 39–47, 2019{\natexlab{b}}.

\bibitem[Zanker and Jessenitschnig(2009)]{DBLP:journals/umuai/ZankerJ09}
M.~Zanker and M.~Jessenitschnig.
\newblock Case-studies on exploiting explicit customer requirements in
  recommender systems.
\newblock \emph{User Modeling and User-Adapted Interaction}, 19\penalty0
  (1-2):\penalty0 133--166, 2009.

\bibitem[Zeng et~al.(2018)Zeng, Nakano, Morita, Kobayashi, and
  Yamaguchi]{zeng2018eliciting}
J.~Zeng, Y.~I. Nakano, T.~Morita, I.~Kobayashi, and T.~Yamaguchi.
\newblock Eliciting user food preferences in terms of taste and texture in
  spoken dialogue systems.
\newblock In \emph{MHFI '18}, page 1–5, 2018.

\bibitem[Zhang and Pu(2006)]{DBLP:conf/ah/ZhangP06}
J.~Zhang and P.~Pu.
\newblock A comparative study of compound critique generation in conversational
  recommender systems.
\newblock In \emph{AH '02}, pages 234--243, June 2006.

\bibitem[Zhang et~al.(2018)Zhang, Chen, Ai, Yang, and
  Croft]{Zhang:2018:TCS:3269206.3271776}
Y.~Zhang, X.~Chen, Q.~Ai, L.~Yang, and W.~B. Croft.
\newblock Towards conversational search and recommendation: {S}ystem ask, user
  respond.
\newblock In \emph{CIKM '18}, pages 177--186, 2018.

\bibitem[Zhao et~al.(2019)Zhao, Fu, Song, Sakai, Chen, Xie, and
  Qian]{zhao2019personalized}
G.~Zhao, H.~Fu, R.~Song, T.~Sakai, Z.~Chen, X.~Xie, and X.~Qian.
\newblock Personalized reason generation for explainable song recommendation.
\newblock \emph{ACM Transactions on Intelligent Systems and Technology},
  10\penalty0 (4):\penalty0 1--21, 2019.

\bibitem[Zhao et~al.(2013)Zhao, Zhang, and Wang]{Zhao:2013:ICF:2541154.2505690}
X.~Zhao, W.~Zhang, and J.~Wang.
\newblock Interactive collaborative filtering.
\newblock In \emph{CIKM '13}, pages 1411--1420, 2013.

\end{thebibliography}



\end{document}